\documentclass{statsoc}
\usepackage{natbib}
\usepackage{graphicx}
\usepackage{amsmath,latexsym}
\usepackage{amsfonts,amssymb}

\newcounter{propietats}
\newenvironment{propietats}
{\begin{list}{\emph{Property} \arabic{propietats}.}%
            {\setlength\labelsep{10pt}%
             \setlength\itemindent{10pt}%
             \setlength\labelwidth{0pt}%
             \setlength\leftmargin{0pt}
             \usecounter{propietats}}}%
{\end{list}}

\title[The normal distribution in some constrained sample spaces]{The normal distribution in some constrained sample spaces}
\author[G. Mateu-Figueras {\it et al.}]{G. Mateu-Figueras}
\address{Universitat de Girona, Girona, Spain.}
\email{gloria.mateu@udg.edu}
\author{V. Pawlowsky-Glahn}
\address{Universitat de Girona, Girona, Spain.}
\author[G. Mateu-Figueras, V. Pawlowsky-Glahn and J.J. Egozcue]{J.J. Egozcue}
\address{Universitat Polit\'ecnica de Catalunya, Barcelona, Spain.}

\begin{document}

\begin{abstract}
Phenomena with a constrained sample space appear frequently in
practice. This is the case e.g. with strictly positive data and
with compositional data, like percentages and the like. If the
natural measure of difference is not the absolute one, it is
possible to use simple algebraic properties to show that it is
more convenient to work with a geometry that is not the usual
Euclidean geometry in real space, and with a measure which is not
the usual Lebesgue measure, leading to alternative models which
better fit the phenomenon under study. The general approach is
presented and illustrated both on the positive real line and on
the $D$-part simplex.\par
\end{abstract}

\section{Introduction}
\label{sec:1} In general, any statistical analysis is performed
assuming data to be realisations of real random vectors whose
density functions are defined with respect to the Lebesgue
measure, which is a natural measure in real space and compatible
with its inner vector space structure. Sometimes, like in the case
of observations measured in percentages, random vectors are
defined on a constrained sample space, $E\subset \mathbb{R}^D$,
and methods and concepts used in real space lead to absurd
results, as it is well known from examples like the spurious
correlations between proportions \citep{Pea1897}. This problem can
be circumvented when $E$ admits a meaningful Euclidean space
structure different from the usual one \citep{PE01}. In fact, if
$E$ is an Euclidean space, a measure $\lambda_E$, compatible with
its structure, is obtained from the Lebesgue measure on
orthonormal coordinates \citep{Eat83,Paw03}. Then, a probability
density function, $f^E$, is defined on $E$ as the
\emph{Radom-Nikod\'ym derivative} of a probability measure
$\mathrm{P}$ with respect to $\lambda_E$. The measure $\lambda_E$
has properties comparable to those of the Lebesgue measure in real
space. Difficulties, arising from the fact that the integral
$\mathrm{P}(A) = \int_{A}f^E(\mathbf{x}) d\lambda_E(\mathbf{x})$
is not an ordinary one, are solved working with coordinates
\citep{Eat83}, and in particular working with coordinates with
respect to an orthonormal basis \citep{Paw03}, as properties that
hold in the space of coordinates transfer directly to the space
$E$. For example, for $f^E$ a density function on $E$, call $f$
the density function of the coordinates, and then the probability
of an event $A\subseteq E$ is computed as $ \mathrm{P}(A)
=\int_{V} f(\mathbf{v}) \; d\lambda(\mathbf{v}), $ where $V$ and
$\mathbf{v}$ are the representation of $A$ and $\mathbf{x}$ in
terms of the orthonormal coordinates chosen, and $\lambda$ is the
Lebesgue measure in the space of coordinates. Using $f$ to compute
any element of the sample space, e.g. the expected value, the
coordinates of this element with respect to the same orthonormal
basis are obtained. The corresponding element in $E$ is then given
by the representation of the element in the basis.

Every one-to-one transformation between a set $E$ and real space
induces a real Euclidean space structure in $E$, with associated
measure $\lambda_E$. Particularly interesting are those
transformations related to the measure of difference between
observations, as evidenced by \cite{Gal1879} when introducing the
logarithmic transformation as a mean to acknowledge Fechner's law,
according to which \emph{perception equals log(stimulus)},
formalised by \cite{McA1879}.

This simple approach has acquired a growing importance in
applications, since it has been recognised that many
\emph{constrained sample spaces}, which are subsets of some real
space---like $\mathbb {R}_+$ or the simplex---can be structured as
Euclidean vector spaces \citep{PE01}. It is important to emphasise
that this approach implies using a measure which is different from
the usual Lebesgue measure. Its advantage is that it opens the
door to alternative statistical models depending not only on the
assumed distribution, but also on the measure which is considered
as appropriate or natural for the studied phenomenon, thus
enhancing interpretation. The idea of using not only the
appropriate space structure, but also to change the measure, is a
powerful tool because it leads to results coherent with the
interpretation of the measure of difference, and because they are
mathematically more straightforward.

\section{Probability densities in Euclidean vector spaces}
\label{sec:densities} Let $E\subseteq\mathbb{R}^D$ be the sample
space for a random vector $\mathbf{X}$, i.e. each realization of
$\mathbf{X}$ is in $E$. Assume that there exists a one-to-one
differenciable mapping $ h: E\rightarrow\mathbb{R}^d$ with $d\leq
D$. This mapping allows to define a Euclidean structure on $E$
just translating the standard properties of $\mathbb{R}^d$ into
$E$. The existence of the mapping $ h$ implies some
characteristics of $E$. An important one in this context is that
$E$ must have some border set so that $ h$ transforms
neighborhoods of this border into neighborhoods of infinity in
$\mathbb{R}^d$. For instance, a sphere in $\mathbb{R}^3$ with a
defined pole can be transformed into $\mathbb{R}^2$, but, if no
pole is defined, this is no longer possible.

The inner sum $\oplus$ and the outer product $\odot$ in $E$ are
defined as $ \mathbf{x}\oplus\mathbf{y}= h^{-1}( h(\mathbf{x})+
h(\mathbf{y}))\ , \ \alpha\odot\mathbf{x}= h^{-1}(\alpha\cdot
h(\mathbf{x}))$, where $\mathbf{x}$, $\mathbf{y}$ are in $E$ and
$\alpha\in\mathbb{R}$. With these definitions $E$ is a vector
space of dimension $d$. The metric structure is induced by the
inner product $ \langle \mathbf{x},\mathbf{y}\rangle_E = \langle
h(\mathbf{x}), h(\mathbf{y})\rangle$, which implies the norm and
the distance $ \| \mathbf{x}\|_E = \|  h(\mathbf{x})\| \ , \
\mathrm{d}_E(\mathbf{x},\mathbf{y})= \mathrm{d}( h(\mathbf{x}),
h(\mathbf{y}))$, thus completing the Euclidean structure of $E$,
based on the inner product, norm and distance in $\mathbb{R}^d$,
denoted as $\langle\cdot,\cdot\rangle$, $\|\cdot\|$,
$\mathrm{d}(\cdot,\cdot)$ respectively. By construction, $
h(\mathbf{x})$ is the vector of coordinates of $\mathbf{x}\in E$.
The coordinates correspond to the orthonormal basis in $E$ given
by the images of the canonical basis in $\mathbb{R}^d$ by $
h^{-1}$. The Lebesgue measure in $\mathbb{R}^d$, $\lambda_d$
induces a measure in $E$, denoted $\lambda_E$, just defining
$\lambda_E( h^{-1}(B))=\lambda_d(B)$, for any Borelian $B$ in
$\mathbb{R}^d$.

In order to define pdf's in $E$, a reference measure is needed.
When $E$ is viewed as a subset of $\mathbb{R}^D$, the Lebesgue
measure, $\lambda_D$, can be eventually used. However, if $d<D$
the random vector $\mathbf{X}$ cannot be absolutely continuous
with respect to $\lambda_D$. Our proposal, and a more natural way
to define a pdf for $\mathbf{X}$, is to start with a pdf for the
(random) coordinates $\mathbf{Y}= h(\mathbf{X})$ in
$\mathbb{R}^d$. Assume that $f_{\mathbf{Y}}$ is the pdf of
$\mathbf{Y}$ with respect to the Lebesgue measure, $\lambda_d$, in
$\mathbb{R}^d$, i.e. $\mathbf{Y}$ is absolutely continuous with
respect to $\lambda_d$ and the pdf is the Radom-Nikod\'ym
derivative $f_{\mathbf{Y}}=dP/d\lambda_d$.

The random vector $\mathbf{X}$ is recovered from $\mathbf{Y}$ as
$\mathbf{X}= h^{-1}(\mathbf{Y})$ but, when $D>d$, $ h^{-1}$ can be
restricted to only $d$ of its components; let $ h_d^{-1}$ be such
a restriction and $\mathbf{X}_d= h_d^{-1}(\mathbf{Y})$. The
inverse mapping is denoted by $ h_d(\mathbf{X}_d)= h(\mathbf{X})$.
This means that more than $d$ components result in a redundant
definition of $\mathbf{X}$. When $D=d$, the restriction of $
h^{-1}$ reduces to the identity $ h^{-1}= h_d^{-1}$.

The pdf of $\mathbf{X}_d$ with respect to the Lebesgue measure in
$\mathbb{R}^d$ is computed using the Jacobian rule
\begin{equation}\label{eq:changedensjacobian}
f_{\mathbf{X}_d}(\mathbf{x}_d)=\frac{dP}{d\lambda_d}(\mathbf{x}_d)=f_{\mathbf{Y}}(
h_d(\mathbf{x}_d)) \cdot \left| \frac{\partial
h_d(\mathbf{x}_d)}{\partial \mathbf{x}_d}   \right| \ ,
\end{equation}
where the last term is the $d$-dimensional Jacobian of $ h_d$.

The next step is to express the pdf with respect to $\lambda_E$,
the natural measure in the sample space $E$. The chain rule for
Radom-Nikod\'ym derivatives implies
\begin{equation}\label{eq:chainrule}
f_{\mathbf{X}_d}^E(\mathbf{x}_d)=\frac{dP}{d\lambda_E}(\mathbf{x}_d)=
\frac{dP}{d\lambda_d}(\mathbf{x}_d)\cdot
\frac{d\lambda_d}{d\lambda_E}(\mathbf{x}_d) \ ,
\end{equation}
and the last derivative is
\begin{equation}\label{eq:jacobianchmeasure}
\frac{d\lambda_d}{d\lambda_E}(\mathbf{x}_d)= \left| \frac{\partial
h_d^{-1}( h_d(\mathbf{x}_d))}{\partial \mathbf{y}}    \right| =
\left| \frac{\partial  h_d(\mathbf{x}_d)}{\partial \mathbf{x}_d}
\right|^{-1}   \ ,
\end{equation}
due to the inverse function theorem. Substituting
(\ref{eq:chainrule}) and (\ref{eq:jacobianchmeasure}) into
(\ref{eq:changedensjacobian}),
\begin{equation}\label{eq:chmeasure}
f_{\mathbf{X}}^E(\mathbf{x})=\frac{dP}{d\lambda_E}(\mathbf{x})=
f_{\mathbf{Y}}( h(\mathbf{x})) \ ,
\end{equation}
where the subscripts $d$ have been suppressed because they only
play a role when computing the Jacobians.

The representation of random variables by pdf's defined with
respect to the measure $\lambda_E$ requires a review of the
moments and other characteristics of the pdf's. Following
\cite{Eat83}, the expectation and variance of $\mathbf{X}$ can be
defined as follows. Let $\mathbf{X}$ be a random variable
supported on $E$ and $h:E\rightarrow \mathbb{R}^d$ the coordinate
function defined on $E$. The expectation in $E$ is
\begin{eqnarray}\label{eq:expectation}
\mathrm{E}^E[\mathbf{X}] & = & \int_E^{\oplus} \mathbf{x} \ f^E_\mathbf{X}(\mathbf{x})\ d\mathbf{x}=h^{-1} \left( \int_{\mathbb{R}^d} \mathbf{y} \ f_{h(\mathbf{X})}(\mathbf{y}) \ d\mathbf{y} \right)\label{eq:EEaton}\\ %
\ &=& h^{-1}\left(\mathrm{E}[h(\mathbf{X})]\right) \label{eq:hEhX} \ , \\ %
\nonumber \end{eqnarray} provided the integrals exist in the
Lebesgue sense. This definition deserves some remarks. The first
integral in (\ref{eq:EEaton}) has been superscripted with $\oplus$
because the involved sum is $\oplus$ for elements in $E$. The
practical way to carry out the integral is to represent the
elements of $E$ using coordinates and to integrate using the pdf
of the coordinates; the result is transformed back into $E$.
Finally, (\ref{eq:hEhX}) summarizes the previous equation using
the standard definition of expectation of the coordinates in
$\mathbb{R}^d$.

Variance involves only real expectations and can be identified
with variance of coordinates. Special attention deserves the
metric variance or total variance \citep{Ait86,PE01}. Assuming the
existence of the integrals, metric variability of $\mathbf{X}$
with respect to a point $\mathbf{z}\in E$ can be defined as
$\mathrm{Var}[\mathbf{X},\mathbf{z}]=\mathrm{E}[\mathrm{d}_E^2(\mathbf{X},\mathbf{z})].$
The minimum metric variability is attained for
$\mathbf{z}=\mathrm{E}^E[\mathbf{X}]$, thus supporting the
definition (\ref{eq:EEaton})--(\ref{eq:hEhX}). The metric variance
is then
\begin{equation}
\mathrm{Var}[\mathbf{X}]=\mathrm{E}[\mathrm{d}_E^2(\mathbf{X},\mathrm{E}^E[\mathbf{X}])]
\ . \label{metricvar}
\end{equation}

The mode of a pdf is normally defined as its maximum value,
although local maxima are normally also called modes. However, the
shape and, particularly, the maximum values depend on the
reference measure taken in the Radom-Nikod\'ym derivatives of the
density. Since the Lebesgue measure in the coordinate space,
$\mathbb{R}^d$, corresponds to the measure $\lambda_E$, the mode
can be defined as
$$
\mathrm{Mode}^E[\mathbf{X}]=\mathrm{argmax}_{\mathbf{x}\in E} \{
f_{\mathbf{X}}^E(\mathbf{x}) \}=
h^{-1}\left(\mathrm{argmax}_{\mathbf{y}\in\mathbb{R}^d}\{
f_{h(\mathbf{X})}(\mathbf{y}) \} \right)\ ,
$$
where the usual remarks on multiple modes or asymptotes are in
order.

\section{The positive real line}
\label{sec:2} The real line, with the ordinary sum and product by scalars, has a
vector space structure. The ordinary inner product and the
Euclidean distance are compatible with these operations. But this
geometry is not suitable for the positive real line. Confront, for
example, some meteorologists with two pairs of samples taken at
two rain gauges, $\{5; 10\}$ and $\{100; 105\}$ in mm, and ask for
the difference; quite probably, in the first case they will say
there was double the total rain in the second gauge compared to
the first, while in the second case they will say it rained a lot
but approximately the same. They are assuming a relative measure
of difference. As a result, the natural measure of difference is
not the usual Euclidean one and the ordinary vector space
structure of $\mathbb{R}$ does not behave suitably. In fact,
problems might appear shifting a positive number (vector) by a
negative real number (vector); or multiplying a positive number
(vector) by an arbitrary real number (scalar), because results can
be outside $\mathbb{R}_+$.

There are two operations, $\oplus$, $\odot$, which induce a vector
space structure in $\mathbb{R}_+$ \citep{PE01}. In fact, given
$x,y\in\mathbb{R}_+$, the internal operation, which plays an
analogous role to addition in $\mathbb{R}$, is the usual product
$x\oplus y= x\cdot y$ and, for $\alpha\in\mathbb{R}$, the external
operation, which plays an analogous role to the product by scalars
in $\mathbb{R}$, is $\alpha\odot x=x^{\alpha}$. An inner product,
compatible with $\oplus$ and $\odot$ is $\langle x,y\rangle_+=\ln
x\cdot \ln y$, which induces a norm, $\|x\|_+=|\ln x |$, and a
distance, $\mathrm{d}_+(x,y)= |\ln y -\ln x |$, and thus the
complete Euclidean space structure in $\mathbb{R}_+$. Since
$\mathbb{R}_+$ is a 1-dimensional vector space there are only two
orthonormal basis: the unit-vector $(e)$ and its inverse element
with respect to the internal operation $(e^{-1})$. From now on the
first option is considered and it will be denoted by $e$. Any
$x\in\mathbb{R}_+$ can be expressed as $x=\ln x \odot e= e^{\ln
x}$ which reveals that $h(x)=\ln x$ is the coordinate of $x$ with
respect to the basis $e$. The measure in $\mathbb{R}_+$ can be
defined so that, for an interval $(a,b)\subset \mathbb{R}_+$,
$\lambda_+(a,b)=\lambda(\ln a,\ln b)=| \ln b-\ln a |$  and
$d\lambda_+/d\lambda =1/x$ \citep{Mat03,Paw03}. Following the
notation in Section \ref{sec:densities}, all these definitions can
be obtained by setting $E=\mathbb{R}_+$, $D=d=1$ and $h(x)=\ln x$.
The generalization to $E=\mathbb{R}_+^D$ is straightforward: for
$\mathbf{x}\in \mathbb{R}_+^D$, the coordinate function can be
defined as $h(\mathbf{x})=\ln(\mathbf{x})$, where the logarithm
applies component-wise.

\subsection{The normal distribution on $\mathbb{R}_+$}
\label{sec:21} Using the algebraic-geometric structure in
$\mathbb{R}_+$ and the measure $\lambda_+$, the normal
distribution on $\mathbb{R}_+$ is defined by \cite{MPM02} through
the density function of orthonormal coordinates.

\vskip 0.5cm

\emph{Definition 1.} Let be $(\Omega,{\cal F},P)$ a probability
space. A random variable $X:\Omega\longrightarrow\mathbb{R}_+$ is
said to have a normal on $\mathbb{R}_+$ distribution with two
parameters $\mu$ and $\sigma^2$, written ${\cal
N}_+(\mu,\sigma^2)$, if its density function is
\begin{equation}
f_{X}^+(x)=\frac{dP}{d\lambda_+}(x)=\frac{1}{\sqrt{2\pi}\sigma}\exp\left
( -\frac{1}{2} \frac{(\ln x-\mu)^2}{\sigma^2}\right ),\qquad
x\in\mathbb{R}_+. \label{fordensinor}
\end{equation}

\begin{figure}[!ht]
 \centering
 \includegraphics[totalheight=1.5in, width=7cm]{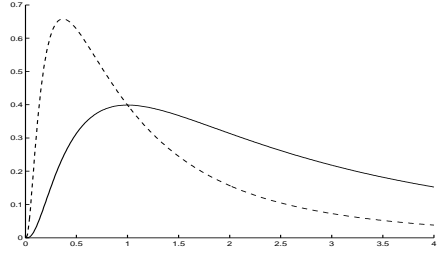}
 \caption{Density functions $\Lambda(0,1)$ (- - - -) and
 ${\cal N}_+(0,1)$ (------).}
 \label{figdensi}
\end{figure}

\noindent The density (\ref{fordensinor}) is the usual normal
density applied to coordinates $\ln x$ as implied by
(\ref{eq:chmeasure}) and it is a density in $\mathbb{R}_+$ with
respect to the $\lambda_+$ measure. This density function is
completely restricted to $\mathbb{R}_+$ and its expression
corresponds to the law of frequency introduced by \cite{McA1879}.
The continuous line in Fig.\ref{figdensi} represents the density
function (\ref{fordensinor}) for $\mu=0$ and $\sigma^2=1$.

According to this approach, the normal distribution in
$\mathbb{R}_+$ exhibits the same characteristics as the normal
distribution in $\mathbb{R}$, the most relevant of which are
summarized in the following properties. A complete proof of the
following properties is presented in the appendix.\par

\begin{propietats}
\item Let be $X\sim {\cal N}_+(\mu,\sigma^2)$, and constants
$a\in\mathbb{R}_+$ and $b\in\mathbb{R}$. Then, the random variable
$X^*= a\oplus(b\odot X)= a\cdot X^{b}$ is distributed as ${\cal
N}_+(\ln a+b\mu, b^2\sigma^2)$.

\item Let be $X\sim {\cal N}_+(\mu,\sigma^2)$ and $a\in\mathbb{R}_+$. %
Then,
$f_{a\oplus X}^+(a\oplus x)=f_{X}^+(x)$,%
where $f_{X}^+$ and $f_{a\oplus X}^+$ represent the probability
density functions of the random variables $X$ and $a\oplus X=
a\cdot X$, respectively.

\item If $X\sim {\cal N}_+(\mu,\sigma^2)$, then
$\mathrm{E}^+[X]=\mathrm{Med}^+[X]=\mathrm{Mode}^+[X]=e^{\mu}$.

\item If $X\sim {\cal N}_+(\mu,\sigma^2)$, then
$\mathrm{Var}[X]=\sigma^2$.
\end{propietats}

Notice that Property 1 implies that the family ${\cal N}_+(\mu,\sigma^2)$ is closed under
the operations in $\mathbb{R}^+$ and Property 2 asserts the
invariance under translations in $\mathbb{R}^+$.

\begin{figure}[!ht]
\begin{tabular}{cc}
\includegraphics[width=5.5cm]{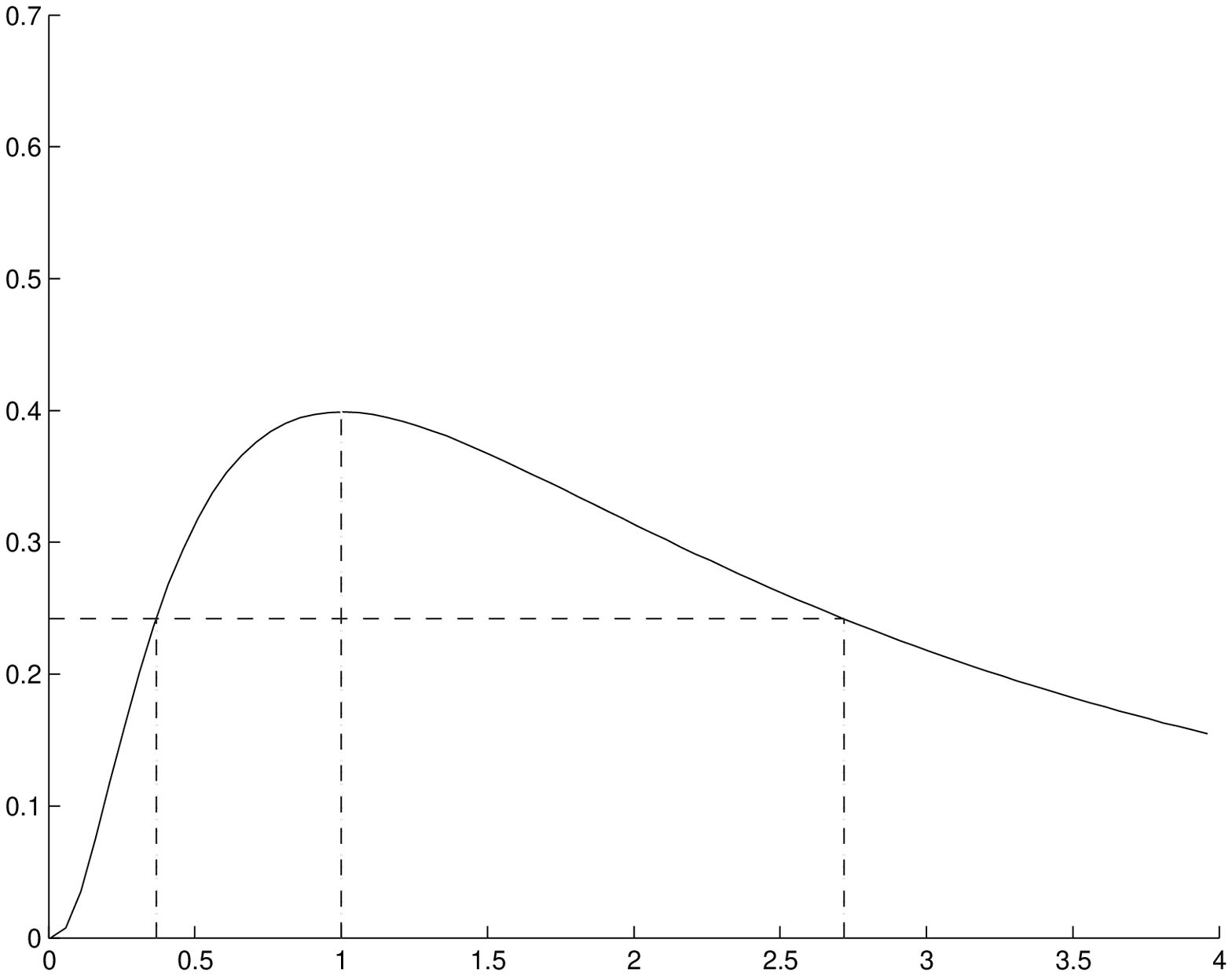} & \includegraphics[width=5.5cm]{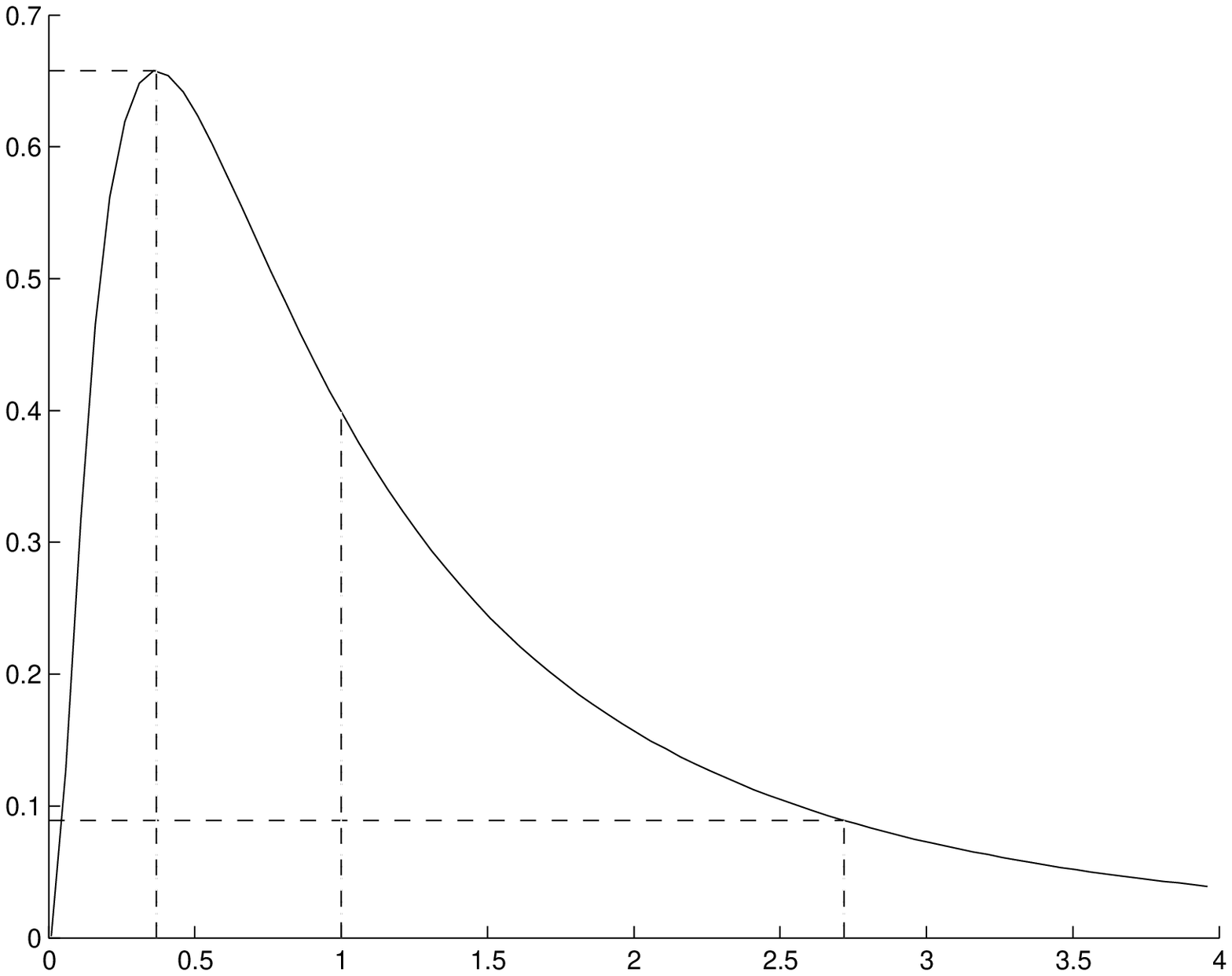}  \\
(a) & (b)
\end{tabular}
 \caption[parida]{\sl Interval $(e^{\mu-\sigma},e^{\mu+\sigma})$ in dashed line (a) ${\cal N}_+(\mu=0,\sigma^2=1)$, (b) $\Lambda(\mu=0,\sigma^2=1)$.}
\label{figint}
\end{figure}

The expected value, the median and the mode are elements of the
support space $\mathbb{R}_+$, but the variance is only a numerical
value which describes the dispersion of $X$. We are used to take
the square root of $\sigma^2$ as a way to represent intervals
centered at the mean and with radius equal to some standard
deviations. To obtain such an interval centered at
$\mathrm{E}[X]=e^{\mu}$ with length $2k\sigma$, take
$(e^{\mu-k\sigma},e^{\mu+k\sigma})$ as
$\mathrm{d}_+(e^{\mu-k\sigma},e^{\mu+k\sigma})=2k\sigma$. This
kind of interval is used in practice \citep{Ahr54} and predictive
intervals in $\mathbb{R}_+$ taking exponential of predictive
intervals computed from the log-transformed data under the
hypothesis of normality are obtained. In Fig.\ref{figint}(a) we
represent the interval $(e^{\mu-\sigma},e^{\mu+\sigma})$ for a
${\cal N}_+(\mu,\sigma^2)$ density function with $\mu=0$ and
$\sigma^2=1$. It can be shown that it is of minimum length, and it
is also an isodensity interval thus, the distribution is
\emph{symmetric} around $e^{\mu}$. This \emph{symmetry} might seem
paradoxical, as one cannot see it in the shape of the density
function. But still, it is symmetric within the linear vector
space structure of $\mathbb{R}_+$, although certainly not within
the Euclidean space structure of $\mathbb{R}_+$ as a subset of
$\mathbb{R}$.

An important aspect of this approach is that consistent estimators
and exact confidence intervals for the expected value are easy to
obtain. We have only to take exponentials of those obtained from
normal theory using log-transformed data, i.e. the coordinates
with respect to the orthonormal basis. Thus, let be
$x_1,x_2,\ldots,x_n$ a random sample and $y_i=\ln x_i$ for
$i=1,2,\ldots,n$, then the optimal estimator for the mean of a
normal in $\mathbb{R}_+$ population is the geometric mean
$(x_1x_2\cdots x_n)^{1/n}$ that equals to $e^{\bar{y}}$. An exact
$(1-\alpha)100\%$ confidence interval for the mean is
$(e^{{\bar{y}}-t_{\alpha/2}V/\sqrt{n}},e^{{\bar{y}}-t_{\alpha/2}V/\sqrt{n}})$
where $V$ denotes the logarithmic variance.

\subsection{Normal on R+ vs lognormal}
\label{sec:22} The lognormal distribution has long been recognized
as a useful model in the evaluation of random phenomena whose
distribution is positive and skew, and specially when dealing with
measurements in which the random errors are multiplicative rather
than additive. The history of this distribution starts in 1879,
when \cite{Gal1879} observed that the law of ``frequency of
errors'' was incorrect in many groups of vital and social
phenomena. This observation was based on Fechner's law which, in
its approximate and simplest form, is ``sensation=log(stimulus)''.
According to this law, an error of the same magnitude in excess or
in deficiency (in the absolute sense) is not equally probable;
therefore, he proposed the geometric mean as a measure of the most
probable value instead of the arithmetic mean. This remark was
followed by the memoir of \cite{McA1879}, where a mathematical
investigation concluding with the lognormal distribution is
performed. He proposed a practical and easy method for the
treatment of a data set grouped around its geometric mean:
``convert the observations into logarithms and treat the
transformed data set as a series round its arithmetic mean'', and
introduced a density function called the ``law of frequency''
which is the normal density function applied to the
log-transformed variable i.e. density (\ref{fordensinor}). In
order to compute probabilities in given intervals, he introduced
also the ``law of facility'', nowadays known as the lognormal
density function.

A unified treatment of the lognormal theory is presented by
\cite{AB57} and more recent developments are compiled by
\cite{CS88}. A great number of authors use the lognormal model
from an applied point of view. Their approach assumes
$\mathbb{R}_+$ to be a subset of the real line with the usual
Euclidean geometry. This is how everybody understands the sentence
``an error of the same magnitude in excess or in deficiency'' in
the same way. One might ask oneself why there is much to say about
the lognormal distribution if the data analysis can be referred to
the intensively studied normal distribution by taking logarithms.
One of the generally accepted reasons is that parameter estimates
are biased if obtained from the inverse transformation.

Recall that a positive random variable $X$ is said to be
lognormally distributed with two parameters $\mu$ and $\sigma^2$
if $Y=\ln X$ is normally distributed with mean $\mu$ and variance
$\sigma^2$. We write $X\sim\Lambda(\mu,\sigma^2)$ and its
probability density function is
\begin{equation}
f_{X}(x)=\left\{\begin{array}{ll}
\frac{1}{\sqrt{2\pi}\sigma x}\exp\left ( -\frac{1}{2}\left(\frac{\ln x-\mu}{\sigma}\right)^2\right ) & x>0,\\
0 & x\leq 0.
\end{array}
\right. \label{fordensilog}
\end{equation}

Comparing (\ref{fordensilog}) with (\ref{fordensinor}), we find
some subtle differences. In fact, the expression of the lognormal
density (\ref{fordensilog}) includes a case for the zero and for
the negative values of the random variable. This fact is
paradoxical, because the lognormal model is completely restricted
to $\mathbb{R}_+$. It is forced by the fact that $\mathbb{R}_+$ is
considered as a subset of $\mathbb{R}$ with the same structure
and, consequently, the variable is assumed to be a real random
variable, hence the name ``lognormal distribution in
$\mathbb{R}$''. Another difference lies in the coefficient $1/x$,
the Jacobian, which is necessary to work with real analysis in
$\mathbb{R}$. More obvious differences are that
(\ref{fordensilog}) is not invariant under translations, that it
is not symmetric around the mean, and that
$\mathrm{E}[X]=e^{\mu+\frac{1}{2}\sigma^2}$, while
$\mathrm{Med}[X]=e^{\mu}$, and both are different from the mode.
The dashed line in Fig.\ref{figdensi} illustrates the probability
density function (\ref{fordensilog}) for $\mu=0$ and $\sigma^2=1$.
Observe that it differs from the density function
(\ref{fordensinor}) plotted in continuous line.\par

However, we can also find some coincidences between the two
models. The median of a $\Lambda(\mu,\sigma^2)$ model is equal to
the median of a ${\cal N}_+(\mu,\sigma^2)$ model. The same happens
with any percentile and any value that involves the distribution
function in its calculation. This property can be easily shown
using measure theory, in particular using properties of
integration with respect to the adequate measure. In fact, given a
lognormal distributed variable $X$ with parameters $\mu$ and
$\sigma^2$, the probability of any interval $(a,b)$ with $0<a<b$
is
$$
P(a<X<b)=\int_{a}^{b}\frac{1}{\sqrt{2\pi}\sigma x}\exp\left (
-\frac{1}{2}\left(\frac{\ln x-\mu}{\sigma}\right)^2\right
)d\lambda(x).
$$

The same probability could be computed using the normal in
$\mathbb{R}_+$ model. Remember that in this case we work in the
coordinates space, thus the probability of any interval $(a,b)$ is
$$
P(a<X<b)=\int_{\ln a}^{\ln b}\frac{1}{\sqrt{2\pi}\sigma}\exp\left
( -\frac{1}{2}\left(\frac{\ln x-\mu}{\sigma}\right)^2\right
)d\lambda(\ln x).
$$

Obviously the same result is obtained in both cases. Therefore we
conclude that the lognormal and the normal in $\mathbb{R}_+$ are
the same probability law over $\mathbb{R}_+$.\par

As we have made for the normal in $\mathbb{R}_+$ case, we could
represent an interval centered at the mean and with radius equal
to some standard deviations for the lognormal in $\mathbb{R}$. If
we consider $\mathbb{R}_+$ as a subset of $\mathbb{R}$ with an
Euclidean structure, these intervals are:
$(\mathrm{E}[X]-k\mathrm{Stdev}[X],\mathrm{E}[X]+k\mathrm{Stdev}[X])$.
But it has no sense, because the lower bound might take a negative
value. For example, for $\mu=0$ and $\sigma^2=1$, the above
interval with $k=1$ is $(-0.512,3.810)$. This is the reason why
sometimes intervals $(e^{\mu-k\sigma},e^{\mu+k\sigma})$ are used,
which are considered to be ``non-optimal'' because they are
neither isodensity intervals, nor do they have minimum length. In
Fig.\ref{figint}(b) we represent the interval
$(e^{\mu-\sigma},e^{\mu+\sigma})$ for the $\Lambda(\mu,\sigma^2)$
density function with $\mu=0$ and $\sigma^2=1$. It is clear that
in the bounds of the interval the density function takes different
values.

Consistent estimators and exact confidence intervals for the mean
and the variance of a lognormal variable are difficult to compute.
Early method of estimating are summarised in \cite{AB57} and
\cite{CS88}. Certainly we find in the literature and extensive
number of procedures and discussions. It is not the objective of
this paper to summarise all methods and to provide a complete set
of formulas. But in general we could say that for the mean, the
term $e^{\bar{y}}$ multiplied by a term expressed as an infinite
serie or tabulated in a set of tables is obtained in most cases
\citep{AB57,Kri81,CH00}. For example, in \cite{CH00} the Sichel's
optimal estimator for the mean of a lognormal population is used.
This estimator is obtained as $e^{\bar{x}}\gamma$, where $\gamma$
is a bias correction factor depending on the variance and the size
of the data set and tabulated in a set of tables. A similar bias
correction factor is used to obtain confidence intervals on the
population mean \citep{CH00}. Nevertheless, in practical
situations, the geometric mean or $e^{\bar{y}}$ is used to
represent the mean and in some cases also to represent the mode of
a lognormal distributed variable \citep{Her60}. But as adverted by
\cite{CS88} those affirmations cannot be justified using the
lognormal theory. On the contrary, using the normal in
$\mathbb{R}_+$ approach those affirmations are completely
justified.

\subsection{Example}
\label{sec:23} The importance of using the normal in
$\mathbb{R}_+$ instead of the lognormal in $\mathbb{R}$ can be
best appreciated in practice.\par

In order to compare the classical lognormal estimators with those
obtained by the normal in $\mathbb{R}_+$ approach, we have
simulated 300 samples representing sizes of oil fields in
thousands of barrels, a geological variable often lognormally
modeled \citep{Dav86}. Using the classical lognormal procedures
and table A2 provided in \cite{AB57} we obtain $161.96$ as an
estimate for the mean. Afterwards and using tables 1,2 and 3 given
in \cite{Kri81} we obtain $162.00$ and $(150.31,176.78)$ as an
estimate and approximate $90\%$ confidence interval for the mean.
Also, using tables 7, 8(b) and 8(e) provided in \cite{CH00} we
could apply the Sichel's bias correction and we obtain $161.86$
and $(144.07,188.39)$ as the optimal estimator and confidence
interval for the mean in the context of the lognormal approach.

Using the normal in $\mathbb{R}_+$ approach we easily obtain
$145.04$ as the estimate for the mean and $(138.70,151.68)$ as the
exact $90\%$ confidence interval for the mean. We have only to
take exponentials of the mean and the $90\%$ confidence interval
obtained from normal theory using log-transformed data. As can be
observed, the differences from those obtained using the lognormal
approach are important. With the normal in $\mathbb{R}_+$ a much
more conservative result is obtained.

\begin{figure}[!ht]
\begin{tabular}{cc}
\includegraphics[width=5.5cm]{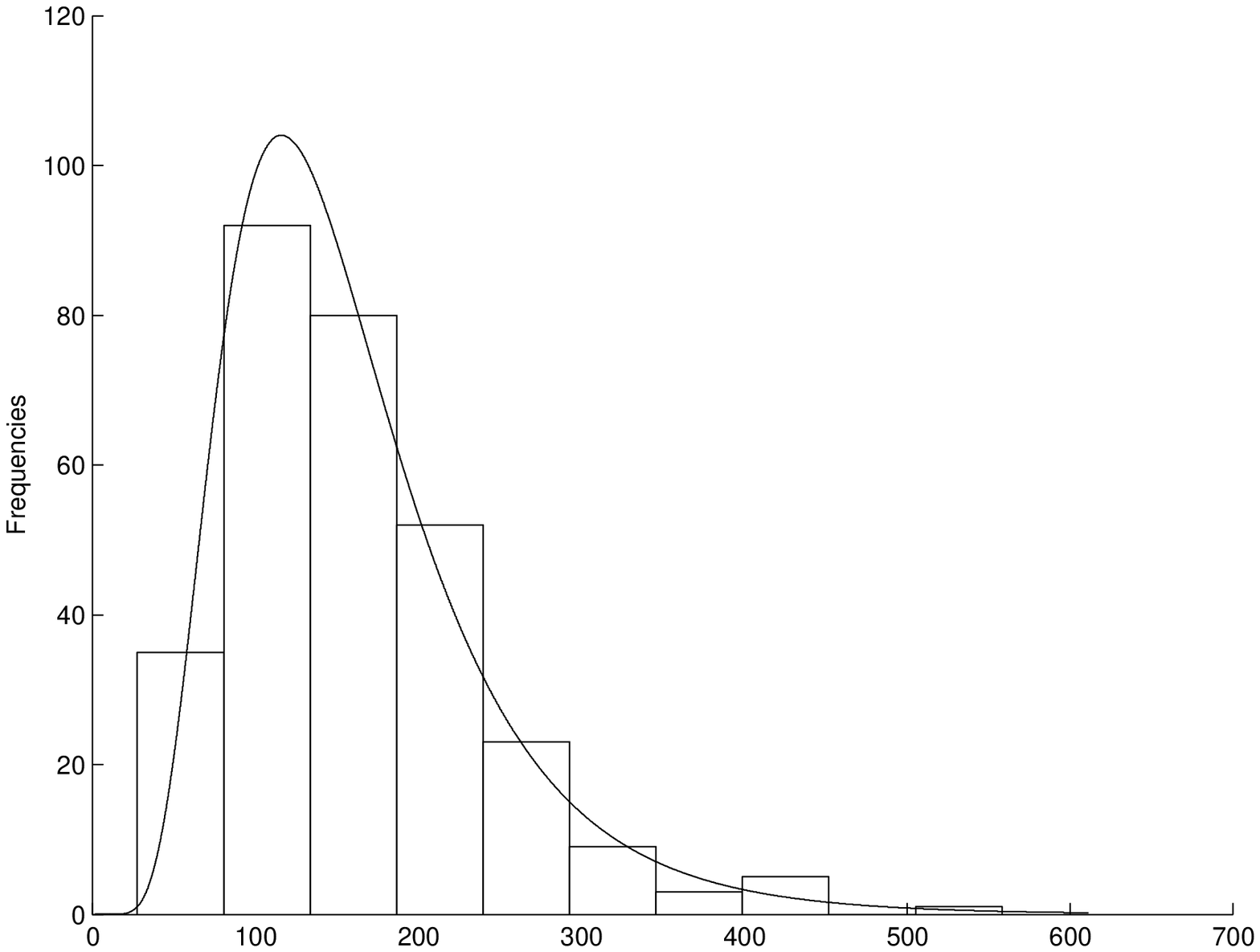} & \includegraphics[width=5.5cm]{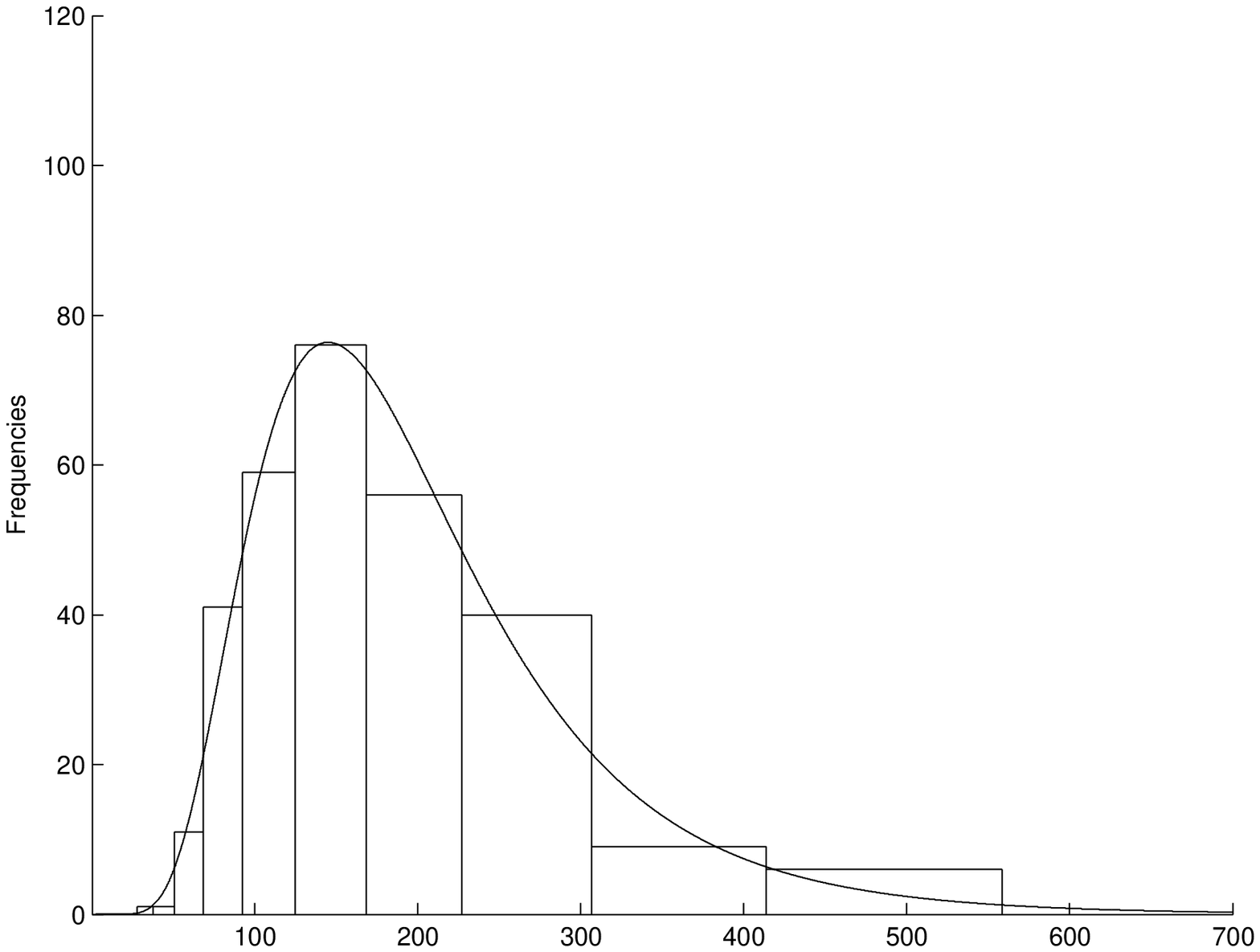}  \\
(a) & (b)
\end{tabular}
 \caption[parida]{Simulated sample $n=300$. Histogram with (a) the fitted lognormal density and
 (b) with the fitted normal in $\mathbb{R}_+$ density.}
\label{fighist}
\end{figure}

\begin{figure}[!ht]
 \centering
 \includegraphics[totalheight=1.5in, width=8cm]{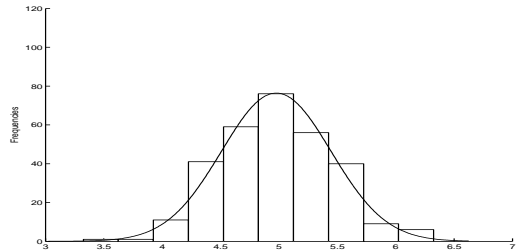}
 \caption[parida]{Simulated sample $n=300$. Histogram of the logtransformed sample with the fitted normal density.}
 \label{fighist2}
\end{figure}

In order to compare graphically the normal in $\mathbb{R}_+$ and
the lognormal approaches we can represent the histogram with the
corresponding fitted densities. In Fig.\ref{fighist}(a) and
\ref{fighist}(b) the histogram with the fitted lognormal and
normal in $\mathbb{R}_+$ densities are provided. Note that the
intervals of the histogram are of equal length in both cases, as
the absolute Euclidean distance is used in (a) and the relative
distance in $\mathbb{R}_+$, $\mathrm{d}_+$, is used in (b) to
compute them. Thus, (b) is a classical histogram but considering
the structure defined in Section 2. Finally, in Fig.\ref{fighist2}
the histogram of the logtransformed data or equivalently of the
coordinates with respect to the orthonormal basis with the fitted
normal density is provided. This last figure is adequate using
both methodologies but in this case we have chosen exactly the
same intervals as in Fig.\ref{fighist}(b). This is only possible
using the normal in $\mathbb{R}_+$ approach because the intervals
on the positive real line have the corresponding intervals in the
space of coordinates.

The normal on $\mathbb{R}_+$ model and its properties has been
recently applied in a spatial context and the results have seen
compared with those obtained with the classical lognormal kriging
approach \citep{TP07}. Using the proposed model and methodology,
the problems of non-optimality, robustness and preservation of
distribution disappear.

\section{The simplex}
\label{sec:3} Compositional data are parts of some whole which
give only relative information. Typical examples are parts per
unit, percentages, ppm, and the like. Their sample space is the
simplex, $ \mathcal{S}^D = \{ \mathbf{x}=(x_1,x_2,\ldots,x_D)':
x_1>0, x_2>0, \ldots, x_D>0; \; \sum_{i=1}^D x_i=\kappa \}$, where
the prime stands for transpose and $\kappa$ is a constant
\citep{Ait82}. For vectors of proportions which do not sum to a
constant, always a fill up value can be obtained.

The simplex $\mathcal{S}^D$ has a $(D-1)$-dimensional Euclidean
space structure \citep{BGF01,PE01} with the following operations.
Let $\mathcal{C}(\cdot)$ denote the closure operation which
normalises any vector $\mathbf{x}$ to a constant sum
\citep{Ait82}, and let be $\mathbf{x}, \mathbf{x}^*
\in\mathcal{S}^D$, and $\alpha\in\mathbb{R}$. Then, the inner sum,
called \emph{perturbation}, is defined as $\mathbf{x} \oplus
{\mathbf{x}^*} = {\cal C}\left( x_1 x_1^*,x_2 x_2^*, \ldots, x_D
x_D^*\right)'$; the outer product, called \emph{powering}, is
defined as $\alpha \odot \mathbf{x} = \mathcal{C} (x_1^{\alpha},
x_2^{\alpha}, \ldots,x_D^{\alpha})'$; and the inner product is
defined as
\begin{equation}
\langle\mathbf{x},\mathbf{x}^*\rangle_a=
\frac{1}{D}\sum_{i<j}\ln\frac{x_i}{x_j}\ln\frac{x_i^*}{x_j^*}\ .
\label{eqs1} \end{equation} The associated squared distance is $
\mathrm{d}_a^2(\mathbf{x},\mathbf{x}^*) = (1/D) \sum_{i<j}
(\ln({x_i}/{x_j}) - \ln({x_i^*}/{x_j^*}))^2. $ This distance is
relative and satisfies standard properties of a distance
\citep{MBP98}, i.e.
$\mathrm{d}_a(\mathbf{x},\mathbf{x}^*)=\mathrm{d}_a(\mathbf{a}\oplus
\mathbf{x},\mathbf{a}\oplus \mathbf{x}^*)$ and
$\mathrm{d}_a(\alpha \odot \mathbf{x},\alpha \odot
\mathbf{x}^*)=\mid\alpha\mid
\mathrm{d}_a(\mathbf{x},\mathbf{x}^*)$. The geometry here defined
is known as \emph{Aitchison geometry}, and therefore the subindex
$a$ is used.

The inner product (\ref{eqs1}) and its associated norm,
$\|\mathbf{x}\|_a = \sqrt{\langle\mathbf{x},\mathbf{x}\rangle_a}$,
ensure the existence of orthonormal basis $\{\mathbf{e}_1,
\mathbf{e}_2, \ldots, \mathbf{e}_{D-1}\}$, which lead to a unique
expression of a composition $\mathbf{x}$ as a linear combination,
$$
\mathbf{x} =
(\langle\mathbf{x},\mathbf{e}_1\rangle_a\odot\mathbf{e}_1) \oplus
(\langle\mathbf{x},\mathbf{e}_2\rangle_a\odot\mathbf{e}_2) \oplus
\ldots \oplus (\langle\mathbf{x},\mathbf{e}_{D-1}\rangle_a \odot
\mathbf{e}_{D-1}).
$$

Like in every inner product space, the orthonormal basis is not
unique. It is not straightforward to determine which one is the
most appropriate to solve a specific problem, but a promising
strategy, based on binary partitions, has been developed by
\cite{EP05}. Here, whenever a specific basis is needed, the basis
given in \cite{EPMB03} is used with respect to which the
coordinates of any $\mathbf{x}\in\mathcal{S}^D$ are
\begin{equation}
y_i=\frac{1}{\sqrt{i(i+1)}}\ln\left(\frac{x_1 x_2\cdots
x_i}{x_{i+1}^i}\right)\ ,\qquad i=1,2,\ldots,D-1\; . \label{eqs3}
\end{equation}
The coordinates in this particular basis are denoted
$\mathrm{ilr}(\mathbf{x})$ to emphasise the similarity with the
vector obtained applying the isometric log-ratio transformation to
a composition $\mathbf{x}$, which is a transformation from
$\mathcal{S}^D$ to $\mathbb{R}^{D-1}$ \citep{EPMB03}. The
important point is that, once an orthonormal basis has been
chosen, all standard statistical methods can be applied to the
coordinates and transferred to the simplex preserving their
properties.

As stated in Section \ref{sec:densities}, the Lebesgue measure in
the space of coordinates induces a measure in $\mathcal{S}^D$,
denoted here as $\lambda_a$. This measure is absolutely continuous
with respect to the Lebesgue measure on real space, and the
relationship between them is $ \left| {d \lambda_a}/{d \lambda}
\right| = (\sqrt{D}\ x_1 x_2 \cdots x_D)^{-1}. $

Following the notation in Section \ref{sec:densities}, all these
definitions can be obtained by setting $E=\mathcal{S}^D$ and
$d=D-1$.

For later use, the concept of subcomposition is required. For $C <
D$, a $C$-part subcomposition, ${\mathbf{x}_S}$, from a $D$-part
composition, $\mathbf{x}$, can be obtained as ${\mathbf{x}_S} =
\mathcal{C}(\mathbf{S} \mathbf{x})$, where $\mathbf{S}$ is a
$C\times D$ selection matrix with $C$ elements equal to 1 (one in
each row and at most one in each column) and the remaining
elements equal to 0 \citep{Ait86}. A subcomposition can be
regarded as a composition in a simplex with fewer parts, and thus
as a space of lower dimension.

\subsection{Some basic statistical concepts in the simplex}
\label{sec:31} A random composition $\mathbf{X}$ is a random
vector with $\mathcal{S}^D$ as domain. In the literature laws of
probability over $\mathcal{S}^D$ can be found, defined using the
standard methodology, i.e. using the Lebesgue measure.
Consequently, the probabilities or any moment are computed using
the classical definition. But some usual elements appear to be of
little use when working with real situations. One typical example
is the expected value which appears as not representative as a
measure of location. As an alternative, the geometric
interpretation of the expected value has been used to define the
centre, $\mathrm{cen}[\mathbf{X}]$, of a random composition as
that composition which minimises the expression $ \mathrm{E}
[\mathrm{d}_a^2 (\mathbf{X} , \mathrm{cen}[\mathbf{X}])]$
\citep{Ait97,PE01}. The result is $\mathrm{cen}[\mathbf{X}] =
\mathcal{C} ( \exp( \mathrm{E}[ \ln\mathbf{X} ] ) )$, which can be
rewritten as \citep{EPMB03} $\mathrm{cen}[\mathbf{X}] =
\mathrm{ilr}^{-1} ( \mathrm{E}[\mathrm{ilr}(\mathbf{X})] )$, or,
in general terms, as
\begin{equation}
\mathrm{cen}[\mathbf{X}] = h^{-1} \left( \mathrm{E}
\left[h(\mathbf{X})\right] \right). \label{eqs4}
\end{equation}

Observe that the centre of a random composition is equal to the
expectation in $\mathcal{S}^D$ defined in Section
\ref{sec:densities}. This is an important result because if a law
of probability on $\mathcal{S}^D$ is defined using the classical
approach, this equality does not hold.

As already mentioned, traditionally the simplex has been
considered as a subset of real space and, consequently, the laws
of probability have been defined using the standard approach. This
is the case for families of distributions like the Dirichlet, the
additive logistic normal \citep{Ait82}, the additive logistic
skew-normal \citep{MPB05}, or those defined using the Box-Cox
family of transformations \citep{Bar96}. Except for the Dirichlet,
these laws of probability are defined using transformations from
the simplex to real space. Two of these transformations will
appear later in this paper, the additive log-ratio
$(\mathrm{alr})$ and the centred log-ratio $(\mathrm{clr})$,
$$
\mathrm{alr}(\mathbf{x})=\left(\ln\left(\frac{x_1}{x_D}\right),
\ldots,\ln\left(\frac{x_{D-1}}{x_D}\right)\right)', \mathrm{
clr}(\mathbf{x})=\left(\ln\left(\frac{x_1}{g(\mathbf{x})}\right),\ldots,
\ln\left(\frac{x_D}{g(\mathbf{x})}\right)\right)',
$$
where $g(\mathbf{x})$ is the geometric mean of composition
$\mathbf{x}$. The relationship between the $\mathrm{alr}$ and the
$\mathrm{clr}$ transformations is provided by \cite{Ait86} (p.92).
The relationships between the $\mathrm{alr}$, $\mathrm{clr}$ and
$\mathrm{ilr}$ transformations are provided by \cite{EPMB03}.

\subsection{The normal distribution on SD}
\label{sec:32} Using the algebraic-geometric structure and the
measure $\lambda_a$ on $\mathcal{S}^D$, the normal distribution on
$\mathcal{S}^D$ is defined through the density function of generic
orthonormal coordinates $h(\mathbf{X})$ \citep{Mat03}. The same
strategy is used in \cite{MP07} to define the skew-normal in
$\mathcal{S}^D$ law.

\vskip 0.5cm

\emph{Definition 2.} Let be $(\Omega,{\cal F},p)$ a probability
space. A random composition
$\mathbf{X}:\Omega\longrightarrow\mathcal{S}^D$ is said to have a
regular normal on $\mathcal{S}^D$ distribution, with parameters
$\mbox{\boldmath \(\mu\)}$ and $\mathbf{\Sigma}$, if its density
function is
\begin{equation}
f_{\mathbf{X}}^{\mathcal{S}}(\mathbf{x}) =(2\pi)^{-(D-1)/2}
|\mathbf{\Sigma}|^{-1/2} \exp \left( -{\frac{1}{2}} \left(
h(\mathbf{x})- \mbox{\boldmath \(\mu\)} \right)'
\mathbf{\Sigma}^{-1} \left( h(\mathbf{x})- \mbox{\boldmath
\(\mu\)} \right)\right), \label{eqs6}
\end{equation}
where $h(.)$ stands for the generic orthonormal coordinates.

\begin{figure}[!ht]
\begin{tabular}{cc}
\includegraphics[width=6.0cm]{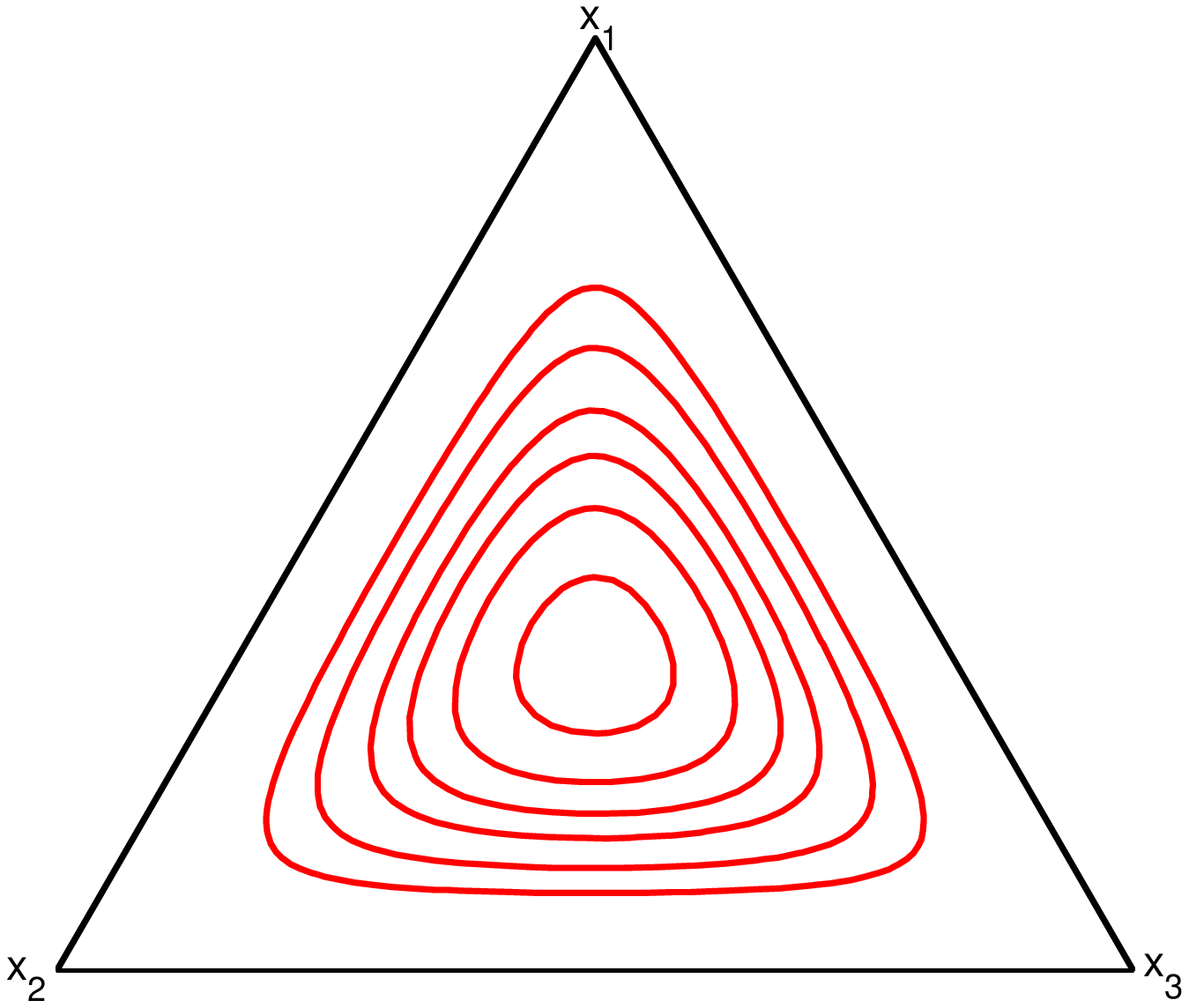} & \includegraphics[width=6.0cm]{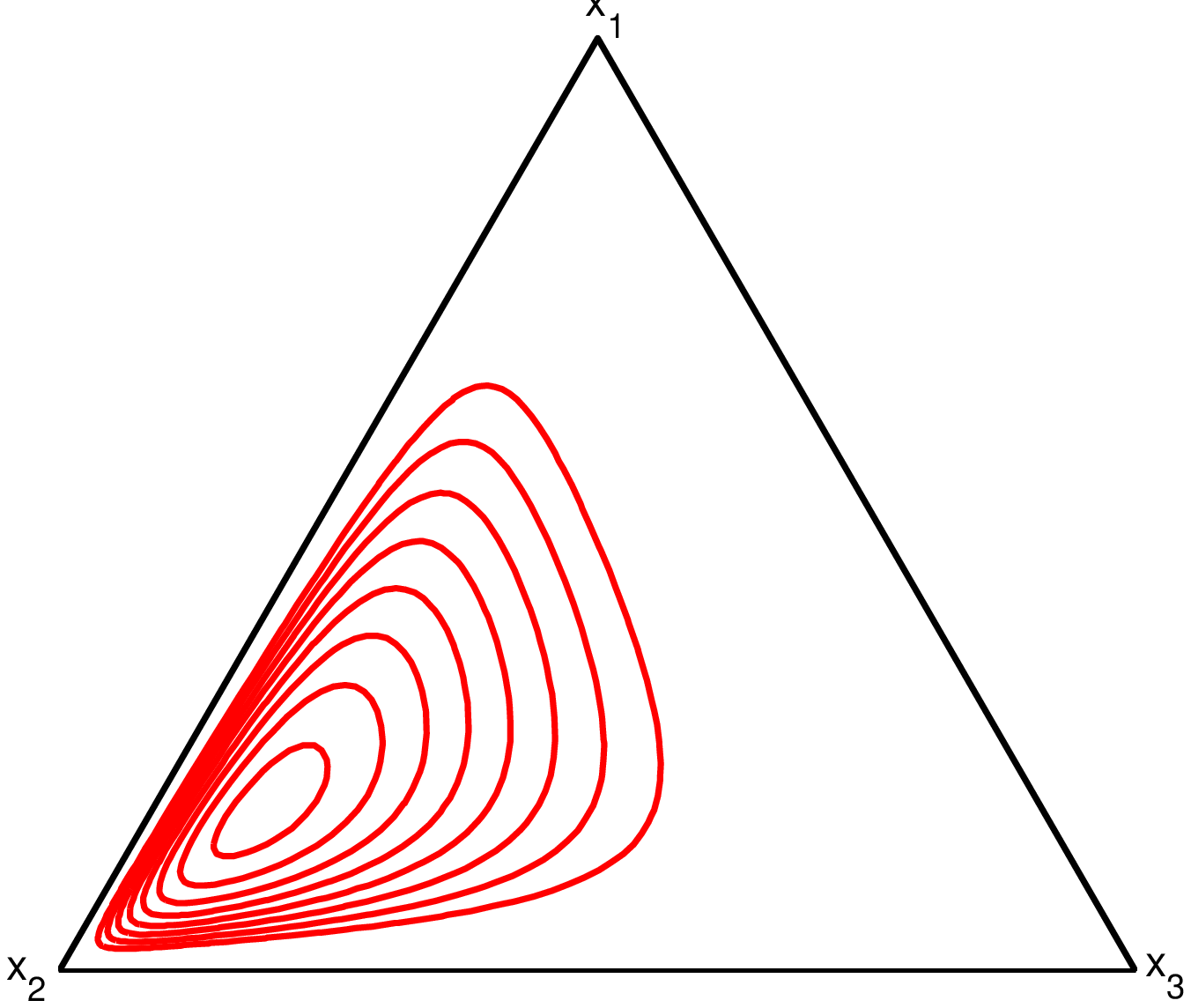}  \\
(a) & (b)
\end{tabular}
 \caption[parida]{Isodensity plots of two $\mathcal{N}_\mathcal{S}^3(\mbox{\boldmath
\(\mu\)},\mathbf{\Sigma})$ with (a) $\mbox{\boldmath
\(\mu\)}=(0,0)$, (b) $\mbox{\boldmath \(\mu\)}=(-1,1)$ and
$\mathbf{\Sigma}=Id$.} \label{fignor}
\end{figure}

The notation $\mathbf{X}\sim
\mathcal{N}_\mathcal{S}^D(\mbox{\boldmath
\(\mu\)},\mathbf{\Sigma},\mbox{\boldmath \(\alpha\)})$ is used.
The subscript $\mathcal{S}$ indicates that it is a model on the
simplex and the superscript $D$ indicates the number of parts of
the composition. Fig.\ref{fignor} shows the isodensity curves of
two normal densities on $\mathcal{S}^3$ taking the particular
basis given by \cite{EPMB03} and using a ternary diagram as a
convenient and simple way for representing 3-part compositions
(see \citealp{Ait86}, p.6).

The density (\ref{eqs6}) is the usual normal density applied to
coordinates $h(\mathbf{x})$ as implied by (\ref{eq:chmeasure}) and
it is a density in $\mathcal{S}^D$ with respect to the $\lambda_a$
measure.

The principal properties of this model follow. A complete proof of
each property can be found in the appendix. The proofs are
straightforward for a reader familiar with compositional data
analysis.

\begin{propietats}
\addtocounter{propietats}{4} \item Let be
$\mathbf{X}\sim\mathcal{N}_\mathcal{S}^D(\mbox{\boldmath
\(\mu\)},\mathbf{\Sigma})$, $\mathbf{a}\in\mathcal{S}^D$ and
$b\in\mathbb{R}$. Then, the $D$-part random composition
$\mathbf{X}^*=\mathbf{a}\oplus(b\odot\mathbf{x})$ has a
$\mathcal{N}_\mathcal{S}^D (h(\mathbf{a})+b\mbox{\boldmath
\(\mu\)},b^2\mathbf{\Sigma})$ distribution.

\item Let be
$\mathbf{X}\sim\mathcal{N}_\mathcal{S}^D(\mbox{\boldmath
\(\mu\)},\mathbf{\Sigma})$ and $\mathbf{a}\in\mathcal{S}^D$. Then
$f_{\mathbf{a}\oplus\mathbf{X}}^{\mathcal{S}}(\mathbf{a}\oplus\mathbf{x})=
f_{\mathbf{X}}^{\mathcal{S}}(\mathbf{x})$, where
$f_{\mathbf{X}}^{\mathcal{S}}$ and
$f_{\mathbf{a}\oplus\mathbf{X}}^{\mathcal{S}}$ represent the
density functions of the random compositions $\mathbf{X}$ and
$\mathbf{a}\oplus\mathbf{X}$, respectively.

\item Let be
$\mathbf{X}\sim\mathcal{SN}_\mathcal{S}^D(\mbox{\boldmath
\(\mu\)},\mathbf{\Sigma})$ and $\mathbf{X}_{P}={\mathbf{P X}}$,
the random composition $\mathbf{X}$ with the parts reordered by a
permutation matrix ${\mathbf{P}}$. Then
$\mathbf{X}_{P}\sim\mathcal{N}_\mathcal{S}^D(\mbox{\boldmath
\(\mu\)}_P,\mathbf{\Sigma}_P)$ with $\mbox{\boldmath
\(\mu\)}_P={\mathbf{U'PU}}\mbox{\boldmath \(\mu\)}$,
$\mathbf{\Sigma}_P=({\mathbf{U'PU}})\mathbf{\Sigma}({\mathbf{U'P'U}})$,
where ${\mathbf{U}}$ is a $D\times(D-1)$ matrix with the clr
transformation of a generic orthonormal basis of $\mathcal{S}^D$
as columns.

\item Let be
$\mathbf{X}\sim\mathcal{N}_\mathcal{S}^D(\mbox{\boldmath
\(\mu\)},\mathbf{\Sigma})$ and
${\mathbf{X}_S}=\mathcal{C}({\mathbf{S}}\mathbf{X})$, the $C$-part
random subcomposition obtained from the $C\times D$ selection
matrix ${\mathbf{S}}$. Then
${\mathbf{X}_S}\sim\mathcal{N}_\mathcal{S}^C(\mbox{\boldmath
\(\mu\)}_S,\mathbf{\Sigma}_S)$, with $\mbox{\boldmath
\(\mu\)}_S={{\mathbf{U}^{*}}}'{\mathbf{SU}}\mbox{\boldmath
\(\mu\)}$,$\mathbf{\Sigma}_S=({\mathbf{U}^{*}}'{\mathbf{SU}})\mathbf{\Sigma}({\mathbf{U'S'}{\mathbf{U}^{*}}})$,
where ${\mathbf{U}}$ is a $D\times(D-1)$ matrix with the
$\mathrm{clr}$ transformation of a generic orthonormal basis of
$\mathcal{S}^D$ as columns and ${\mathbf{U}}^*$ is a
$C\times(C-1)$ matrix with the $\mathrm{clr}$ transformation of a
generic orthonormal basis of $\mathcal{S}^C$ as columns.

\item Let be
$\mathbf{X}\sim\mathcal{N}_\mathcal{S}^D(\mbox{\boldmath
\(\mu\)},\mathbf{\Sigma})$. Then, the expected value in
$\mathcal{S}^D$ is
\begin{equation}
\mathrm{E}_a[\mathbf{X}]=(\mu_1\odot\mathbf{e}_1)
\oplus(\mu_2\odot\mathbf{e}_2)\oplus\ldots\oplus(\mu_{D-1}\odot\mathbf{e}_{D-1}).
\label{eqs7}
\end{equation}

\item Let be
$\mathbf{X}\sim\mathcal{N}_\mathcal{S}^D(\mbox{\boldmath
\(\mu\)},\mathbf{\Sigma})$. The metric variance of $\mathbf{X}$ is
$\mathrm{
Var}[\mathbf{X}]=\hbox{trace}\left(\mathbf{\Sigma}\right)$.
\end{propietats}

From Property 5 we conclude that the normal on $\mathcal{S}^D$ law
is closed under perturbation and powering. From Property 6 we see
that it is also invariant under perturbation. This has important
consequences, because when working with compositional data the
centring operation \citep{MBBP99}, a perturbation using the
inverse of the centre of the data set, is often applied in
practice to better visualise and interpret the pattern of
variability \citep{EPE02}.

Notice that Properties 7 and 8 show that the normal on $\mathcal{S}^D$ family
is closed under permutation and subcompositions.

Given a compositional data set the estimates of parameters
$\mbox{\boldmath \(\mu\)}$ and $\mathbf{\Sigma}$ can be computed
applying the maximum likelihood procedure to the coordinates of
the data set. The estimated values $\hat{\mbox{\boldmath
\(\mu\)}}$ and $\widehat{\mathbf{\Sigma}}$ allow us to compute the
estimates of the expected value and metric variance of random
composition $\mathbf{X}$, as $
\widehat{\mathrm{E}_a[\mathbf{X}]}=(\hat{\mu}_1\odot\mathbf{e}_1)
\oplus\cdots\oplus(\hat{\mu}_{D-1}\odot\mathbf{e}_{D-1})$ and
$\widehat{\mathrm{
Var}[\mathbf{X}]}=\hbox{trace}\left(\widehat{\mathbf{\Sigma}}\right)$.

To validate the distributional assumption of normality on
$\mathcal{S}^D$, some goodness-of-fit tests of the multivariate
normal distribution have to be applied to the coordinates of the
sample data set. There is a large battery of possible tests but as
suggested by \cite{Ait86} we could start testing the normality of
each marginal using empirical distribution function tests.
Unfortunately, the univariate normality of each component is a
necessary but not sufficient condition for the normality of the
whole vector. Also, these univariate tests depend on the
orthonormal basis chosen. This difficulty does not depend on the
proposed methodology, as the same problem appears when working
with laws of probability defined using transformations and the
Lebesgue measure in $\mathcal{S}^D$ \citep{AMN03}. The
multivariate normal model can also be validated considering the
Mahalanobis distance $( h(\mathbf{X})- \hat{\mbox{\boldmath
\(\mu\)}})' \widehat{\mathbf{\Sigma}}^{-1} ( h(\mathbf{X})-
\hat{\mbox{\boldmath \(\mu\)}})$ which is sampled from a
$\chi^2_{D-1}$-distribution if the fitted model is appropriate. In
this case, the dependence on the chosen orthonormal basis
disappears. Here, the use of empirical distribution function tests
is also suggested \citep{Ait86}.

\subsection{The normal on SD vs the additive logistic normal}
\label{sec:33} The classical approach is used by \cite{Ait82} to
define the additive logistic normal law on the simplex. The
strategy is standard: transform the random composition from the
simplex to the real space, define the density function of the
transformed vector and finally go back to the simplex using the
theorem of the change of variable. The result is a density
function for the initial random composition with respect to the
Lebesgue measure. Thus, a random composition is said to have an
additive logistic normal distribution ($\mathrm{aln}$) when the
additive log-ratio transformed vector has a normal distribution.
Note that this definition does not explicitly state that the
theorem of the change of variable has to be used. But this is the
principal difference between this approach, based on working with
transformations, with the new approach, based on working with
coordinates.

The $\mathrm{aln}$ model was initially defined using the additive
log-ratio transformation. Using the matrix relationship among the
log-ratio transformations \citep{EPMB03} we can easily obtain the
density function in terms of the isometric log-ratio
transformation. Thus we can define the logistic normal
distribution with parameters $\mbox{\boldmath \(\mu\)}$ and
$\mathbf{\Sigma}$, with density function:
\begin{equation}
f_{\mathbf{X}}(\mathbf{x})
=\frac{(2\pi)^{-(D-1)/2}\mid\mathbf{\Sigma}\mid^{-1/2}}{\sqrt{D}x_1
x_2 \cdots x_D}\exp \left( -{\frac{1}{2}}\left(
\mathrm{ilr}(\mathbf{x})-\mbox{\boldmath
\(\mu\)}\right)'\mathbf{\Sigma}^{-1}
\left(\mathrm{ilr}(\mathbf{x})-\mbox{\boldmath
\(\mu\)}\right)\right).\label{eqs9}
\end{equation}

\begin{figure}[!ht]
\begin{tabular}{cc}
\includegraphics[width=6.0cm]{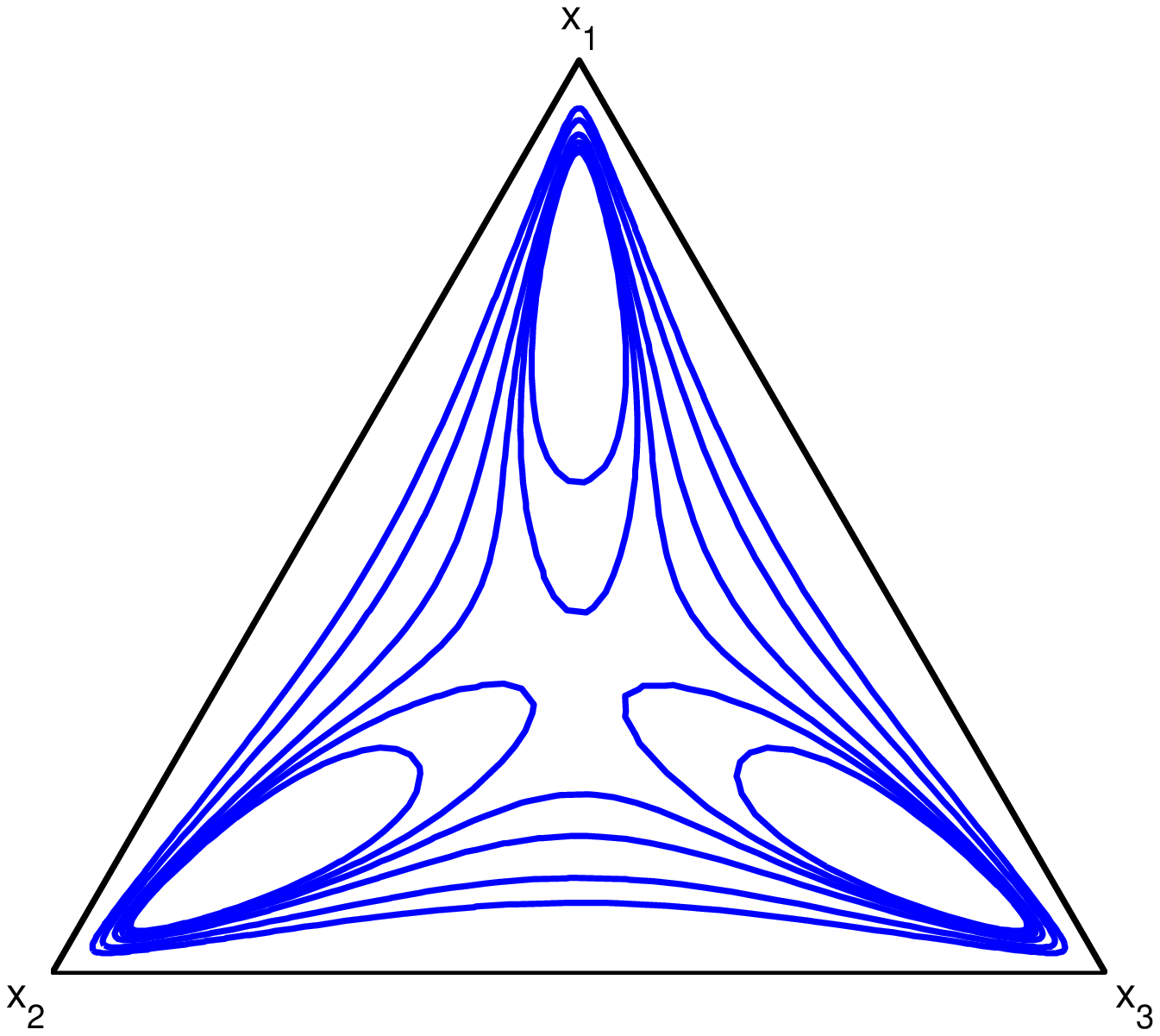} & \includegraphics[width=6.0cm]{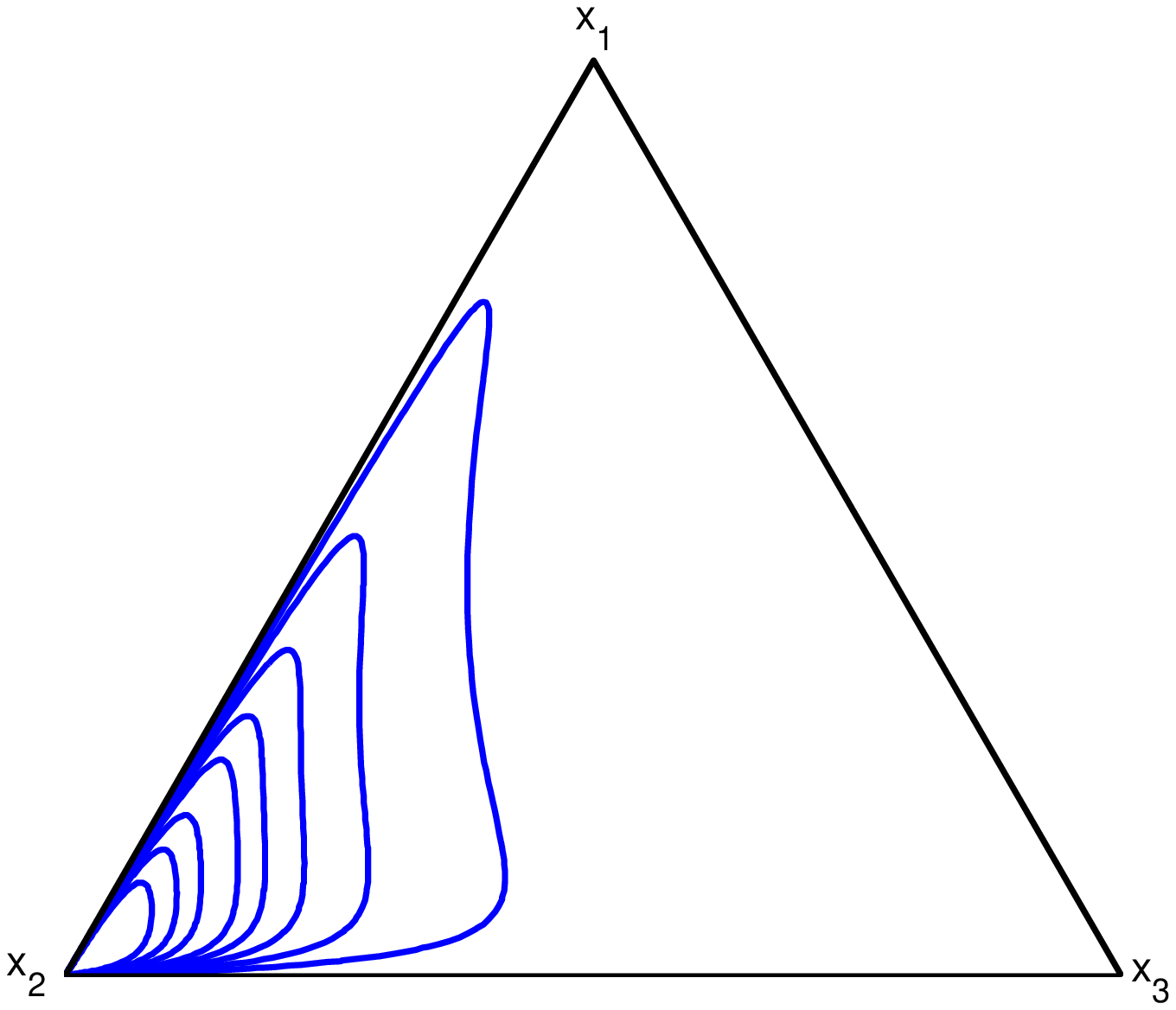}  \\
(a) & (b)
\end{tabular}
 \caption[parida]{Isodensity plots of two logistic normal models with (a) $\mbox{\boldmath
\(\mu\)}=(0,0)$, (b) $\mbox{\boldmath \(\mu\)}=(-1,1)$ and
$\mathbf{\Sigma}=Id$.} \label{figaln}
\end{figure}

To easily compare both approaches we will use the normal model on
the simplex taking the basis given in \cite{EPMB03} and
consequently the $\mathrm{ilr}$ vector stated in (\ref{eqs3}).
Nevertheless, we could consider any orthonormal basis as we can
obtain vector $\mathrm{ilr}(\mathbf{x})$ from $h(\mathbf{x})$ and
the corresponding change of basis matrix. If we compare the
expression of the densities (\ref{eqs6}) and (\ref{eqs9}), the
only difference is the term $(\sqrt{D} x_1 x_2 \cdots x_D)^{-1}$,
the jacobian of the isometric log-ratio transformation that
reflects the change of the measure on $\mathcal{S}^D$. The
influence of this term can be observed in the isodensity curves in
Fig.\ref{figaln}. These curves can be directly compared with the
curves in Fig.\ref{fignor}. The differences are obvious, in
particular the trimodality in Fig.\ref{figaln}(a). This behaviour
is not exclusive of the logistic normal model, we find also
bimodality with Beta or Dirichlet densities when their parameters
tend to 0 and when the Lebesgue measure is considered. In
Fig.\ref{figaln}(b) we observe a unique mode, nevertheless its
position and the shape of the curves are not the same as in
Fig.\ref{fignor}(b), the corresponding normal on
$\mathcal{S}^3$.

Another essential difference between the two models are the
moments of any order. We know that the expression of the density
function plays a fundamental role when any moment is computed. The
density (\ref{eqs9}) is a classical density, consequently we
compute any moment using the standard definition. Obviously the
results are not the same as in the normal on $\mathcal{S}^D$ case.
For example, the expected value of an $\mathrm{aln}$ model exists,
but numerical procedures have to be applied (see \citealp{Ait86},
p.116) to find it and the result is not the same as in Property 9.

Also, some coincidences can be found.  The closure under
perturbation, powering, permutation and subcompositions of the
logistic normal model is proved by \cite{Ait86}, the same as those
stated in Properties 5,7 and 8 for the normal on $\mathcal{S}^ D$
model. Nevertheless the logistic normal class is not invariant
under perturbation, that is,
$f_{\mathrm{a}\oplus\mathbf{X}}(\mathrm{a}\oplus\mathbf{x})\neq
f_{\mathbf{X}}(\mathbf{x})$.

Another coincidence is that the two models assign the same
probability to the events and we can say that both models are
equivalent on $\mathcal{S}^D$. In fact, given a logistic normal
distributed random composition $\mathbf{X}$ with parameters
$\mbox{\boldmath \(\mu\)}$ and $\mathbf{\Sigma}$, the probability
of any event $A\subseteq\mathcal{S}^D$ is
\begin{equation}
P(A)=\int_{A}\frac{(\sqrt{D} x_1 x_2 \cdots
x_D)^{-1}}{\mid\mathbf{\Sigma}\mid^{1/2}(2\pi)^{(D-1)/2}}\exp
\left( -{\frac{1}{2}}\left(
\mathrm{ilr}(\mathbf{x})-\mbox{\boldmath
\(\mu\)}\right)'\mathbf{\Sigma}^{-1}\left(\mathrm{ilr}(\mathbf{x})-\mbox{\boldmath
\(\mu\)}\right)\right)d\lambda(\mathbf{x}), \label{eqs10}
\end{equation}
where now the vector $\mathrm{ilr}(\mathbf{x})$ denotes the
isometric log-ratio transformation of vector $\mathbf{x}$. The
same probability using the normal on $\mathcal{S}^D$ model is
\begin{equation}
P(A)=\int_{\mathrm{ilr}(A)}\frac{1}{\mid\mathbf{\Sigma}\mid^{1/2}(2\pi)^{(D-1)/2}}\exp
\left( -{\frac{1}{2}} \left(\mathbf{v}-\mbox{\boldmath
\(\mu\)}\right)' \mathbf{\Sigma}^{-1}
\left(\mathbf{v}-\mbox{\boldmath
\(\mu\)}\right)\right)d\lambda(\mathbf{v}), \label{eqs12}
\end{equation}
with $\mathrm{ilr}(A)$ giving now the representation of event $A$
in coordinates with respect to the particular orthonormal basis
given by \cite{EPMB03}. At this point it is important to correctly
interpret the vector $\mathrm{ilr}(\mathbf{x})$ as the isometric
log-ratio transformed vector or as the vector of coordinates.
Therefore, to avoid possible confusions, we denote by $\mathbf{v}$
the vector of coordinates in expression (\ref{eqs12}). Certainly,
the two vectors are numerically identical, but here the meaning is
important.

Both expressions (\ref{eqs10}) and (\ref{eqs12}) are standard
integrals of a real valued function. Thus, we can apply a change
of variable in (\ref{eqs12}), taking
$\mathbf{v}=\mathrm{ilr}(\mathbf{x})$ whose jacobian is $(\sqrt{D}
x_1 x_2 \cdots x_D)^{-1}$, and the equality
$$
P(A)=\int_{A}\frac{(\sqrt{D} x_1 x_2 \cdots
x_D)^{-1}}{\mid\mathbf{\Sigma}\mid^{1/2}(2\pi)^{(D-1)/2}}\exp
\left(-{\frac{1}{2}}\left(
\mathrm{ilr}(\mathbf{x})-\mbox{\boldmath
\(\mu\)}\right)'\mathbf{\Sigma}^{-1}\left(\mathrm{ilr}(\mathbf{x})-\mbox{\boldmath
\(\mu\)}\right) \right)d\lambda(\mathbf{x})
$$
is obtained. This equality agrees with (\ref{eqs10}) given that
$\mathrm{ilr}^{-1}(\mathrm{il r}(A))=A$. Remember that an
isometric log-ratio transformed element is equal to its
coordinates with respect to the orthonormal basis given by
\cite{EPMB03}. Then, the $\mathrm{ilr}^{-1}$ transformation gives
the original element on the simplex. Therefore we conclude that
the additive logistic normal law and the normal on $\mathcal{S}^D$
law are the same probability law over the simplex.

Concerning estimation and goodness-of-fit testing, we will obtain
exactly the same results using both models. Remember that in the
normal in $\mathcal{S}^D$ case we work with the $\mathrm{ilr}$
coordinates whereas in the logistic normal case we work with the
$\mathrm{ilr}$ transformed vector.

In summary, the essential differences between both approaches are
the shape of the probability density function, in some cases
leading to multimodality for the standard approach; the moments
which characterise the density, particularly important in practice
for the expected value and the variance; and invariance under
perturbation.

\subsection{Example}
\label{sec:34} To illustrate the differences between using a
density with respect to the Lebesgue measure $\lambda$ or a
density with respect to the measure $\lambda_a$ in
$\mathcal{S}^D$, the Skye lavas data \citep{TED72} will be used.
It contains chemical compositions of 23 basalt specimens from the
Isle of Skye in the form of percentages of 10 major oxides. This
data set is used in \cite{Ait82} to discuss the adequacy of some
parametric models and no significant indication of non-normality
is obtained for the $\mathrm{alr}$ transformed data set. Due to
the matrix relationship between the $\mathrm{alr}$ and
$\mathrm{ilr}$ transformations, we can easily conclude no
significant departure from normality for the $\mathrm{ilr}$
transformed data set.

Our objective in this section is to compare graphically the
logistic normal and the normal on the simplex. Thus, in order to
provide some useful figures a 3-part compositional data set is
preferred. For this reason we take $\mathbf{X}$ as the popular AFM
subcomposition (A: $Na_2O+K_2O$, F: $Fe_2O_3$ and M:$MgO$) from
the Skye lavas data set. The resulting data can be found in
\cite{Ait86} or in \cite{TED72}. As the first component is
obtained amalgamating two original parts, we cannot guarantee the
adequacy of the logistic normal and the normal in $\mathcal{S}^3$
models. Following the suggestions by \cite{Ait86} we could test
the goodness-of-fit of the model applying a battery of 12 tests,
based on the Anderson-Darling, Cram\'{e}r-von Mises and Watson
statistics, to the coordinates of the sample data set. In
particular, the tests are applied to the marginal distributions,
to the bivariate angle distribution and to the radius. Taking a 1
per cent significance level only one of the marginal tests gives
evidence of any departure from normality.

The fit of a normal model on $\mathcal{S}^3$ and of a logistic
normal model (using the $\mathrm{ilr}$ transformation) gives, as
noted in the previous section, exactly the same estimates of the
parameters for both models:
$$
\hat{\mbox{\boldmath \(\mu\)}}=(0.555,0.639)',\qquad
\mathbf{\hat{\Sigma}}=\left(\begin{array}{cc} 0.126 & -0.229 \\
                                   -0.229 & 0.456 \\
                                   \end{array}\right).
$$
Here, the orthonormal basis given by \cite{EPMB03} has been used,
and consequently the $\mathrm{ilr}$ vector stated in (\ref{eqs3}).

\begin{figure}[!ht]
\begin{tabular}{cc}
\includegraphics[width=6cm]{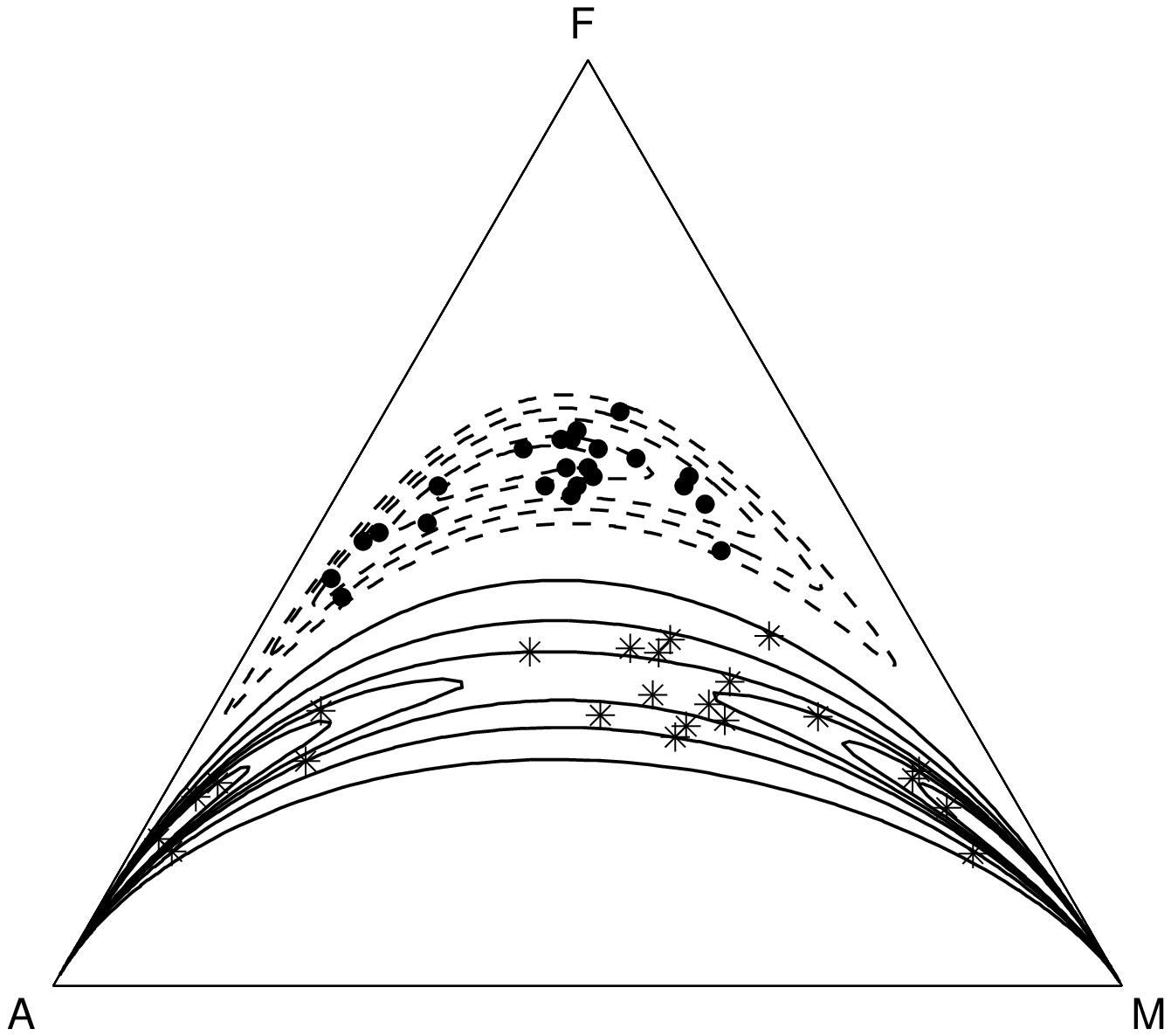} & \includegraphics[width=6cm]{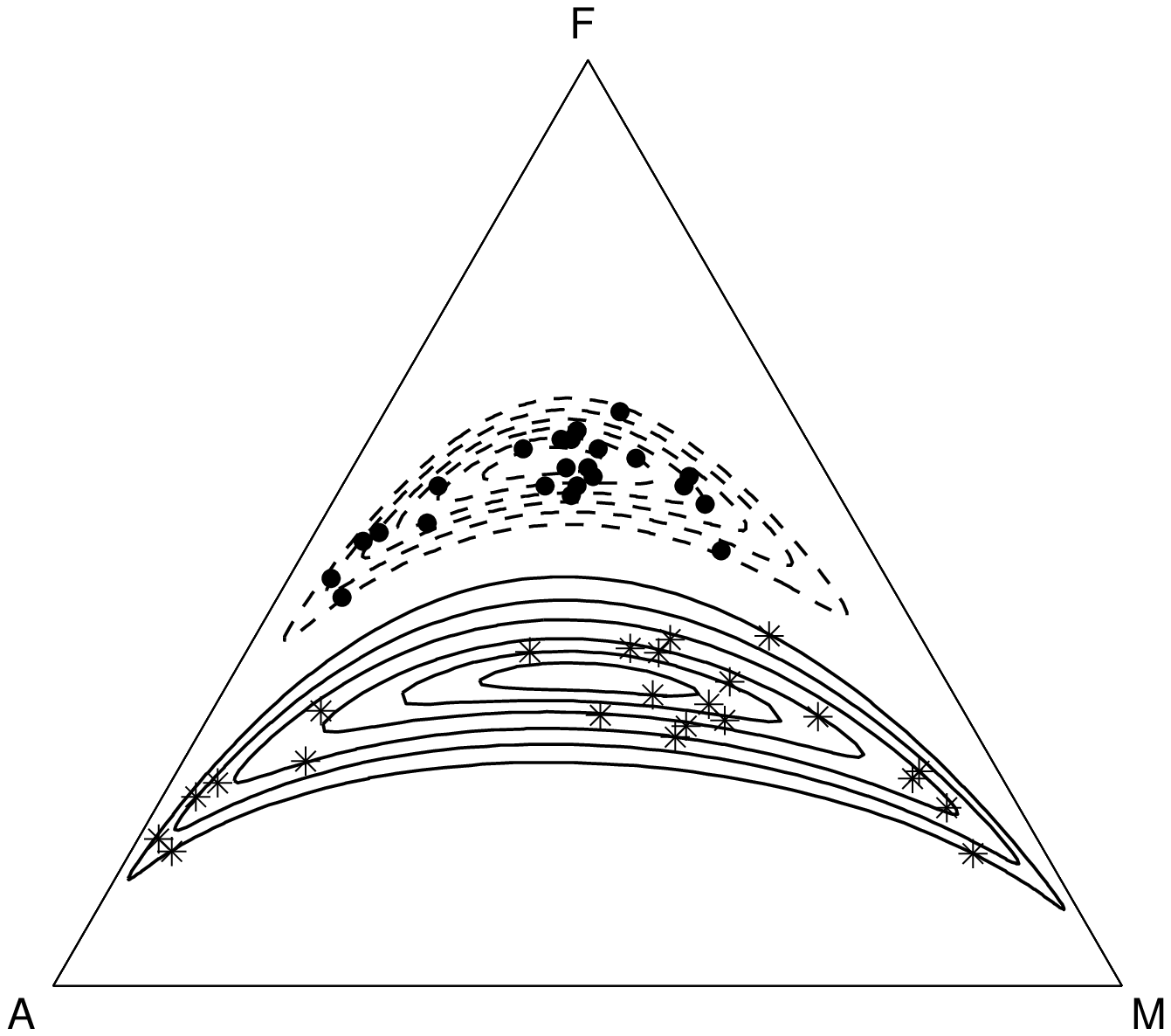}  \\
(a) & (b)
\end{tabular}
 \caption[parida]{\footnotesize\sl Skye lavas data $(\bullet)$ and
 linearly transformed data $(\ast)$ with isodensity curves of the fitted (a)
 logistic normal model and (b) normal on $\mathcal{S}^3$.}
\label{figlavas}
\end{figure}

The fit of the logistic normal and normal in $\mathcal{S}^3$
models are represented in dashed line in Fig.\ref{figlavas}(a) and
\ref{figlavas}(b) respectively. The two fitted models are quite
similar.

As both models follow Property 5, i.e. the families are
closed under perturbation and powering, the transformation
$\mathbf{a}\oplus(b\odot\mathbf{X})$ is applied to the data, with
$\mathbf{a}=g(\mathbf{X})^{-1}$ and $b=\sqrt{3}$. This is a linear
transformation in $\mathcal{S}^3$ and has been chosen only for
illustration purposes. Note that the geometric mean of our
resulting data set is the center of the simplex, composition
$(1/3,1/3,1/3)$, because we first modify the variability applying
the power operation but then we center our data. It is equivalent
to translate the transformed data set, or the coordinates with
respect to an orthonormal basis, to the origin of coordinates in
the real space. For both resulting models the estimates of the
parameters follow the equations stated in Property 5 i.e.
$$
\hat{\mbox{\boldmath \(\mu\)}}=(0.000,0.000)',\qquad
\mathbf{\hat{\Sigma}}=\left(\begin{array}{cc} 0.377 & -0.688 \\
                                   -0.688 & 1.369 \\
                                   \end{array}\right).
$$

\begin{figure}[!ht]
 \centering
 \includegraphics[totalheight=1.8in]{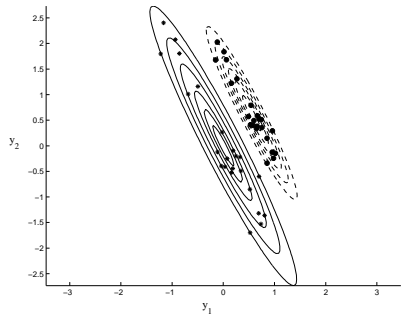}
 \caption[parida]{\sl $\mathrm{ilr}$ coordinates of the Skye lavas data set and the corresponding fitted normal models
 to the original data (dashed line) and to the linear transformed data
 (continuous line).}
 \label{figlavasilr}
\end{figure}

In Fig.\ref{figlavas}(a) and \ref{figlavas}(b) the logistic normal
and the normal in $\mathcal{S}^3$ fitted models are represented in
continuous line. As can be observed, the same linear
transformation leads to a better visualisation of the normal on
$\mathcal{S}^3$ fitted model, but in the logistic normal case a
completely different model, with two modes, is obtained. In other
words, perturbation and powering, which should only move the
centre of the density and modify the variability, can generate
arbitrary modes, an undesirable property. In Fig.\ref{figlavasilr}
we represent the corresponding normal densities fitted to the
$\mathrm{ilr}$ coordinates or equivalently to the $\mathrm{ilr}$
transformed data set, because the same graphic is obtained using
both methodologies. It is clear that the linear transformation
only increase the variability and translate our data set to the
origin of coordinates.

\section{Conclusions}
A particular Euclidean vector space structure of the positive real
line and of the simplex, together with the associated measure,
allow us to define parametric models with desirable properties.
Normal models on $\mathbb{R}_+$ and on $\mathcal{S}^D$ have been
defined through their densities over the coordinates with respect
to an orthonormal basis and their main algebraic properties have
been studied. From a probabilistic point of view, those laws of
probability are identical to the lognormal and to the additive
logistic normal distribution defined using the Lebesgue measure
and the standard methodology based on transformations.
Nevertheless, some differences are obtained in the moments and in
the shape of the density function. In particular, the expected
value differs from what would be obtained with the lognormal and
with the additive logistic normal distributions, something
important when they are used to characterise real data using a
probabilistic model.


\section*{Acknowledgments}
This work has been supported by the Spanish Ministry of Education
and Science under project 'Ingenio Mathematica (i-MATH)' No.
CSD2006-00032 (Consolider – Ingenio 2010) and under project
MTM2006-03040.



\bibliographystyle{natbib}

\vskip 2mm

\noindent APPENDIX

\noindent This appendix contains the proofs of properties
contained in Sections \ref{sec:21} and \ref{sec:32}. The
construction of these proofs is done using the expected value, the
covariance matrix, the linear transformation property of the
multivariate normal distribution and some matrix relationships
among vectors of coordinates and among log-ratio
transformations.\par \vskip 3mm

\noindent \emph{Proof of Property 1.} The coordinates of the
random variable $X^*$ are obtained from the coordinates of the
variable $X$ as $\ln(X^*) = \ln(a) + b\ln (X)$. The density
function of $\ln(X)$ is the classical normal density in the real
line; thus, the linear transformation property can be used to
obtain the density function of the $\ln(X^*)$ random variable.
Therefore, $X^*\sim {\cal N}_+(\ln a+b\mu, b^2\sigma^2)$.\par

\vskip 2mm

\noindent \emph{Proof of Property 2.} From Property 1 we know that
$a\oplus X= a\cdot X\sim {\cal N}_+(\ln a+\mu, \sigma^2)$. From
(\ref{fordensinor}) we get
$$
f^+_{ a\oplus X}(a\oplus x) =
\frac{1}{\sqrt{2\pi}\sigma}\exp\left( -\frac{1}{2} \frac{(\ln
(ax)-(\ln a+\mu))^2 }{\sigma^2}\right )= f^+_{X}(x).
$$

\vskip 2mm

\noindent \emph{Proof of Property 3.} From (\ref{eq:hEhX}) we
known that $\mathrm{E}^E[X]=\exp((\mathrm{E}[\ln X])$, and from
(\ref{fordensinor}) we known that the density function of $\ln X$
is the normal distribution, thus $\mathrm{E}^E[X]=\exp(\mu)$. The
same result is obtained for the median and the mode as the normal
distribution is symmetric around its expected value $\mu$.

\vskip 2mm

\noindent \emph{Proof of Property 4.} From (\ref{metricvar}) we
know that the variance can be understood as the expected value of
the squared distance around its expected
value, i.e. $\mathrm{Var}[X]= \\
\mathrm{E}[\mathrm{d}_+^2(X,\mathrm{E}^+[X])]$. Working on
coordinates and using the density function of $\ln X$ we obtain
$\mathrm{Var}[X]=\mathrm{E}[\mathrm{d}^2(\ln{X},\mathrm{E}[\ln{X}])]=
\mathrm{Var}[\ln{X}]=\sigma^2$.

\vskip 2mm

\noindent \emph{Proof of Property 5.} The orthonormal coordinates
of the random composition $\mathbf{X}^*$ are obtained from the
orthonormal coordinates of the composition $\mathbf{X}$ via
$h(\mathbf{X}^*) = h(\mathbf{a}) + b h(\mathbf{X})$. The density
function of $h(\mathbf{X})$ random vector is the classical normal
density in real space; thus, the linear transformation property
can be used to obtain the density function of the
$h(\mathbf{X}^*)$ random vector. Therefore,
$\mathbf{X}^*\sim\mathcal{N}_\mathcal{S}^{D}(h(\mathbf{a})+b\mbox{\boldmath
\(\mu\)},b^2\mathbf{\Sigma},\mbox{\boldmath \(\alpha\)})$.\par

\vskip 2mm

\noindent \emph{Proof of Property 6.} Using Property 5,
$\mathbf{a}\oplus\mathbf{X}\sim\mathcal{N}_\mathcal{S}^D(h(\mathbf{a})+\mbox{\boldmath
\(\mu\)},\mathbf{\Sigma},\mbox{\boldmath \(\alpha\)})$. We known
that $h(\mathbf{a}\oplus\mathbf{x})=h(\mathbf{a})+h(\mathbf{x})$,
therefore,
\begin{align*}
f_{\mathbf{a}\oplus\mathbf{X}}(\mathbf{a}\oplus\mathbf{x}) &=
(2\pi)^{-(D-1)/2}\mid\mathbf{\Sigma}\mid^{-1/2}\\
&\ \times\exp \left[ -{\frac{1}{2}}\left(
h(\mathbf{a}\oplus\mathbf{x})-(h(\mathbf{a})+\mbox{\boldmath
\(\mu\)})\right)'\mathbf{\Sigma}
^{-1}\left(h(\mathbf{a}\oplus\mathbf{x})-(h(\mathbf{a})+\mbox{\boldmath
\(\mu\)})\right)\right]=\\
&=(2\pi)^{-(D-1)/2}\mid\mathbf{\Sigma}\mid^{-1/2}\exp \left[
-{\frac{1}{2}}\left( h(\mathbf{x})-\mbox{\boldmath
\(\mu\)}\right)'\mathbf{\Sigma}
^{-1}\left(h(\mathbf{x})-\mbox{\boldmath
\(\mu\)}\right)\right]=f_{\mathbf{X}}(\mathbf{x}).
\end{align*}\par

\vskip 2mm

\noindent \emph{Proof of Property 7.} For the centered log-ratio
transformed vectors it is straightforward to see that $\mathrm{
clr}(\mathbf{X}_P)={\mathbf{P}}\mathrm{ clr}(\mathbf{X})$ \cite[p.
94]{Ait86}. Using the matrix relationship between the centered and
the isometric log-ratio vectors \citep{EPMB03} we conclude that
$h(\mathbf{X}_P)=({\mathbf{U'PU}})h(\mathbf{X})$. Given the
density of the $h(\mathbf{X})$ random vector, and applying the
linear transformation property of the normal distribution in real
space, a $\mathcal{N}_\mathcal{S}^D(\mbox{\boldmath
\(\mu\)}_P,\mathbf{\Sigma}_P,\mbox{\boldmath \(\alpha\)}_P)$
distribution is obtained for the random composition
$\mathbf{X}_{P}$.\par

\vskip 2mm

\noindent \emph{Proof of Property 8.} \cite[p. 119]{Ait86} gives
the matrix relationship between $\mathrm{ alr}(\mathbf{X}_S)$ and
$\mathrm{ alr}(\mathbf{X})$. Using the matrix relationships
between the additive, centered and isometric log-ratio vectors
\citep{EPMB03}, we conclude that
$h(\mathbf{X}_S)=({\mathbf{U}^{*}}'{\mathbf{ SU}})h(\mathbf{X})$.
Given the density of the $h(\mathbf{X})$ vector, and applying the
linear transformation property of the normal distribution in real
space, the density of the $h(\mathbf{X}_S)$ vector is obtained as
that of the $\mathcal{N}_\mathcal{S}^C(\mbox{\boldmath
\(\mu\)}_S,\mathbf{\Sigma}_S,\mbox{\boldmath \(\alpha\)}_S)$
distribution. \par

\vskip 2mm

\noindent \emph{Proof of Property 9.} From (\ref{eq:hEhX}) we
known that $\mathrm{E}_a[\mathbf{X}]=
h^{-1}(\mathrm{E}[h(\mathbf{X})])$, and from Definition 2
we know that the density function of $h(\mathbf{X})$ is the
multivariate normal distribution; thus
$\mathrm{E}[h(\mathbf{X})]=\mbox{\boldmath \(\mu\)}$. Finally, the
composition $\mathrm{E}_a[\mathbf{X}]$ is obtained applying
$h^{-1}$ or by the representation of this element in the basis
$(\mu_1\odot\mathbf{e}_1)\oplus
(\mu_2\odot\mathbf{e}_2)\oplus\ldots
\oplus(\mu_{D-1}\odot\mathbf{e}_{D-1})$.\par

\vskip 2mm

\noindent \emph{Proof of Property 10.} From (\ref{metricvar}) we
know that the variance can be understood as the expected value of
the squared distance around its expected value, i.e.
$\mathrm{Var}[\mathbf{X}]=\mathrm{E}[d_a^2(\mathbf{X},\mathrm{E}_a[\mathbf{X}])]$.
Working on coordinates and using the density function of
$h(\mathbf{X})$ we obtain $\mathrm{ Var}[\mathbf{X}]=
\mathrm{trace}(\mathbf{\Sigma})$.\par

\end{document}